\documentclass[11pt,a4paper]{article}
\usepackage{times}
\usepackage{dsfont}
\usepackage{fullpage}
\usepackage{titlesec}
\usepackage[bookmarks=false]{hyperref}
\usepackage[normalem]{ulem}
\usepackage{graphicx,cite}
\usepackage{geometry}
\usepackage{amssymb,amsmath,amsthm}
\usepackage{natbib}
\usepackage{amssymb,bm}
\usepackage{array}
\usepackage{amsfonts}
\usepackage{enumitem}
\usepackage{color}

\usepackage{geometry,array,colortbl,xcolor}
\usepackage{tabulary}
\usepackage{xcolor}
\usepackage{colortbl}
\usepackage[normalem]{ulem} 
\usepackage{multirow}

\usepackage{bbding}
\usepackage{pifont}
\usepackage{wasysym}
\usepackage{amssymb}
\usepackage{subcaption}
\usepackage{ifthen}

\usepackage{setspace}
\usepackage{amsmath,amsfonts}
\usepackage{amssymb}
\usepackage{graphicx}

\usepackage{amsmath,amscd,amssymb,amsgen,amsfonts,amsbsy,amsthm}
\usepackage{mathrsfs}
\usepackage{bm}
\usepackage{bbm}
\usepackage[utf8x]{inputenc}
\usepackage{color}
\usepackage{textcomp}
\usepackage{float}
\usepackage{latexsym,graphicx}
\usepackage{url}
\usepackage{todonotes}
\usepackage{authblk}

\theoremstyle{plain}
\newtheorem{definition}{Definition}

\newtheorem{theorem}{Theorem}
\newtheorem{lemma}[theorem]{Lemma}
\newtheorem{corollary}[theorem]{Corollary}
\newtheorem{proposition}[theorem]{Proposition}

\theoremstyle{remark}

\newtheorem{remark}{Remark}

\newcommand{\graya}[1]{ #1} 
\makeatletter
\def\highlight#1{%
	\fboxrule1pt %
	\hsize=\dimexpr\hsize-3\fboxrule-3\fboxsep\relax
	\@endpbox\unskip\setbox\lastbox\bgroup
	\fboxrule1pt %
	\fcolorbox{black}{white}{\box{ #1}}\hfill}

\def\Cline#1#2{\@Cline#1#2\@nil}
\def\@Cline#1-#2#3\@nil{%
	\omit
	\@multicnt#1%
	\advance\@multispan\m@ne
	\ifnum\@multicnt=\@ne\@firstofone{&\omit}\fi
	\@multicnt#2%
	\advance\@multicnt-#1%
	\advance\@multispan\@ne
	\leaders\hrule\@height#3\hfill
	\cr}
\newcommand\Thickvrule[1]{%
	\multicolumn{1}{c!{\vrule width 1pt}}{#1}%
}

\begin{document}
\vspace*{3ex}
\begin{center}
	\LARGE{Improving the Performance of Heterogeneous Data Centers through Redundancy}
	\vspace*{3ex}
	
	\large
	Elene Anton$^{1,3}$, Urtzi Ayesta$^{1,2,3,4}$, Matthieu Jonckheere$^{5}$ and Ina Maria Verloop$^{1,3}$
\end{center}
\vspace*{1ex}

\begin{center}
	\small
	\textsuperscript{1} CNRS, IRIT, 31071 Toulouse, France.\\
	\textsuperscript{2} {IKERBASQUE - Basque Foundation for Science, 48011 Bilbao, Spain.}\\
	\textsuperscript{3} {Universit\'e de Toulouse, INP, 31071 Toulouse, France.}\\
	\textsuperscript{4} {UPV/EHU, University of the Basque Country, 20018 Donostia, Spain.}\\
	\textsuperscript{5} {Instituto de C\'alculo - Conicet, Facultad de Ciencias Exactas y Naturales,} \\
	Universidad de Buenos Aires, 1428 Buenos Aires, Argentina. \\
	\normalsize
\end{center}
\vspace*{1ex}

\begin{abstract}
	We analyze the performance of redundancy in a multi-type job and multi-type server system. We assume the job dispatcher is unaware of the servers' capacities, and we set out to study under which circumstances redundancy improves the performance. With redundancy an arriving job dispatches redundant copies to all its compatible servers, and departs as soon as one of its copies completes service. As a benchmark comparison, we take the non-redundant system in which a job arrival is routed to only one randomly selected compatible server. Service times are generally distributed and all copies of a job are identical, i.e., have the same service requirement.
	
	In our first main result, we characterize the sufficient and necessary stability conditions of the redundancy system. This condition coincides with that of a system where each job type only dispatches copies into its least-loaded servers, and those copies need to be  fully served.
	In our second result, we compare the stability regions of the system under redundancy to that of no redundancy. 
	We show that if the server's capacities are sufficiently heterogeneous, the stability region under redundancy can be much larger than that without redundancy. We apply the general solution to particular classes of systems, including redundancy-$d$ and nested models, to derive simple conditions on the degree of heterogeneity required for redundancy to improve the stability. As such, our result is the first in showing that redundancy can  improve the stability and hence performance of a system when copies are  \emph{non-i.i.d.}.
\end{abstract}

\noindent\textit{Key words:} redundancy models; load balancing; stochastic stability; processor sharing.

\section{Introduction}
\label{sec:intro}

The main motivation of studying redundancy models comes from the fact that both empirical  (\cite{Ananthanarayanan10,Ananthanarayanan13,Dean13,Vulimiri13})
and theoretical (\cite{Joshi15,Shah16,Gardner16,Lee17a,Lee17b,Gardner17b}) evidence show that redundancy might improve the performance of real-world applications. 
Under redundancy, a job that arrives to the system dispatches multiple copies into the servers, and departs when a first copy completes service. 
By allowing for redundant copies, the aim is to minimize the latency of the system by exploiting the variability in  the queue lengths and the capacity of the different servers.


Most of the theoretical results on redundancy systems consider the performance analysis when either FCFS or Processor-Sharing (PS) service policies are implemented in the servers. Under the assumption that all the copies of a job are i.i.d. (independent and identically distributed) and exponentially distributed, \cite{Gardner16,Bonald17a,Anton2019} show that the stability condition of the system is independent of the number of redundant copies and that performance (in terms of delay and number of jobs in the system) improves as the number of copies increases. However, \cite{Gardner17b} showed that the assumption that copies of a job are i.i.d. \ can be unrealistic, and that it might lead to theoretical results that do not reflect the results of replication schemes in real-life computer systems. The latter has triggered interest to consider other modeling assumptions for the correlation structure of the copies of a job. For example, for identical copies (all the copies of a job have the same size), \cite{Anton2019} showed that under both FCFS and PS service policies, the stability region of the system with \emph{homogeneous} servers decreases as the number of copies increases. 

The above observation provides the motivation for our study: to understand when redundancy is beneficial. 
In order to do so, we analyze a general multi-type job and multi-type server system. A dispatcher needs to decide to which server(s) to route each incoming job.
We assume that there is no signaling between the dispatcher and the servers, that is, the dispatcher is oblivious to the capacities of the servers and unaware of the states of the queues. The latter can be motivated by \emph{(i)} design constraints, \emph{(ii)} (slowly) fluctuating capacity of a server due to external users, or \emph{(iii)} the impossibility of exchanging information among dispatchers and servers. The only information that is available to the dispatcher is the type of job and its set of compatible servers. However, we do allow signaling \textit{between/among servers}, which is needed in order to cancel the copies in redundancy schemes.

In the mathematical analysis  we consider two different models: the redundancy model where the dispatcher sends a copy to all the compatible servers of the job type, and
the Bernoulli model where a single copy is send to a uniformly selected compatible server of the job type. From a dispatchers viewpoint, the comparison between these two policies is reasonable under the assumption that the dispatcher only knows the type of the job and the set of its compatible servers. Hence, we do not compare analytically the performance of redundancy with other routing policies -- such as Join the Shortest Queue, Join the Idle Server, Power of $d$, etc. -- that have more information on the state of the system. 
We 
hence aim to understand when having redundant copies is beneficial for the performance of the system in this context. Observe that the answer is not clear upfront as adding redundant copies has two opposite effects: on the one hand, redundancy helps exploiting the variability across servers' capacities, but on the other hand, it induces a waste of resources as servers work on copies that do not end up being completely served. 

To answer the above question, we analyze the stability of an arbitrary multi-type job and multi-type server system with redundancy. Job service requirements are generally distributed, and copies are identical. 
The scheduling discipline implemented by servers is PS, which is a common policy in server farms and web servers, see for example \cite[Chapter 24]{HB13}. In our main result, we derive sufficient and necessary stability conditions for the redundancy system. 
This general result allows us to characterize when redundancy can increase the stability region with respect to Bernoulli routing. 

To the best of our knowledge, our analytical results are the first showing that, when copies are non-i.i.d., adding redundancy to the system can be beneficial from the stability point of view.  We believe that our result can motivate further research in order to thoroughly understand when redundancy is beneficial in other settings. For example, for different scheduling disciplines, different correlation structures among copies, different redundancy schemes, etc. In Section~\ref{sec:numerics} we investigate through numerics some of these issues, namely, the performance of redundancy when the scheduling discipline is FCFS and Random Order of Service (ROS), and the performance gap between redundancy and a variant of Join the Shortest Queue policy according to which each job is dispatched to the compatible server that has the least number of jobs.

We briefly summarize the main findings of the paper:
\begin{itemize}
	\item The characterization of sufficient and necessary stability conditions of any general redundancy system with heterogeneous server capacities and arrivals, under mild assumptions on the service time distribution. 
	\item 
	We prove that when servers are heterogeneous enough (conditions stated in Section~\ref{sec:improve}), redundancy has a larger stability region than Bernoulli.
	\item By exploring numerically these conditions, we observe that the degree of heterogeneity needed in the servers for redundancy to be better, decreases  in the number of servers, and increases in the number of redundant copies.

\end{itemize}

The rest of the paper is organized as follows. 
In Section~\ref{sec:related} we discuss related work.  
Section~\ref{sec:model} describes the model, and introduces the notion of \emph{capacity-to-fraction-of-arrivals ratio} that plays a key role in the stability result. Section~\ref{sec:example}  gives an illustrative example in order to obtain intuition about the structure of the stability conditions. 
Section~\ref{sec:gen_red} states the stability condition for the redundancy model. Section~\ref{sec:improve} provides conditions on the heterogeneity of the system under which redundancy outperforms Bernoulli. 
The proof of the main result is given in Section~\ref{sec:stab_cond_proof}. Simulations are given in Section~\ref{sec:numerics}, and concluding remarks are given in Section~\ref{sec:conclusion}. For the sake of readability,   proofs are deferred to the Appendix.

\section{Related work} \label{sec:related}

When copies of a job are i.i.d. and exponentially distributed,  \cite{Gardner16,Bonald17a} have shown that redundancy with FCFS employed in the servers does not reduce the stability region of the system. In this case, the stability condition is that for any subset of job types, the sum of the arrival rates must be smaller than the sum of service rates associated with these job types. 
In~\cite{Raaijmakers2018}, the authors consider i.i.d.\ copies with highly variable service time distributions. They  focus on redundancy-$d$ systems where each job chooses a subset of $d$ homogeneous servers uniformly at random. The authors show that with FCFS, the stability region increases (without bound) in both the number of copies, $d$, and in the parameter that describes the variability in service times.

In \cite{KR08}, the authors investigate when it is optimal to replicate a job. They show that for so-called New-Worse-Than-Used service time distributions, the best policy is to replicate as much as possible. In \cite{GHR19}, the authors investigate the impact that scheduling policies have on the performance of so-called nested redundancy systems with i.i.d.\ copies. The authors show that when FCFS is implemented, the performance might not improve as the number of redundant copies increases, while under other policies proposed in the paper, such as Least-redundant-first or Primaries-first, the performance improves as the number of copies increases.

Anton et al. \cite{Anton2019} study the stability conditions when the scheduling policies  PS, Random Order of Service (ROS) or FCFS are implemented.  For the redundancy-$d$ model with homogeneous server capacities and i.i.d. copies, they 
show that the stability region is not reduced if either PS or Random Order of Service (ROS) is implemented. 
When instead copies belonging to one job are identical, \cite{Anton2019} showed that \emph{(i)} ROS does not reduce the stability region, \emph{(ii)} FCFS reduces the stability region 
and \emph{(iii)} PS dramatically reduces the stability region, and this coincides with the stability region of a system where all copies need to be \emph{fully} served, i.e., $\lambda < \frac{\mu K}{d}$. In \cite{Raaijmakers2019}, the authors show that the stability result for PS in a homogeneous redundancy-$d$ system with identical copies extends to generally distributed service times.
In the present paper, we extend \cite{Anton2019,Raaijmakers2019} by characterizing the stability condition under  PS with identical copies to the general setting of  heterogeneous servers, generally distributed service times, and arbitrary redundancy structures.	

Hellemans et al. \cite{HvH18b} consider identical copies that are generally distributed. For a redundancy-$d$ model with FCFS, they develop a numerical method to compute the workload and response time distribution when the number of servers tends to infinity, i.e., the mean-field regime. The authors can numerically infer whether the system is stable, but do not provide any characterization of the stability region. 
In a recent paper, Hellemans et al. \cite{HBvH20} extend this study to   include many replication policies, and  general correlation structure among the copies.

Gardner at al. \cite{Gardner17b} introduce a new dependency structure among the copies of a job, the S\&X model. The service time of each copy of a job is decoupled into two components: one related to the inherent job size of the task, that is identical for all the copies of a job, and the other one related to the server's slowdown, which is independent among all copies. The paper proposes and analyzes the redundant-to-idle-queue scheme with homogeneous servers, and proves that it is stable, and performs well. 

In Table 1 we summarize the stability results presented above, organized by service policy, service time distribution, servers'  capacities and redundancy correlation structure. In brackets we specify the additional assumptions that the authors considered in their respective paper. In the bold square, we outline the modeling assumptions we consider for the present paper. To the best of our knowledge, no analytical results were obtained so far for performance measures when PS is implemented, servers are heterogeneous \emph{and} copies are identical or of any other non i.i.d. structure. 

\begin{table*}
	\centering
	\caption{The stability condition of redundancy models under different modeling assumptions. In bold square, the modeling assumptions we consider for the present paper.}
	\footnotesize
	\setlength{\tabcolsep}{2pt}
	\renewcommand{\arraystretch}{0.9}	
	\begin{tabular}{|l|l|l|l|l|l|}
		\hline
		\multirow{2}{*}{} & \textbf{Service time} &
		\multicolumn{2}{c|}{\textbf{Homogeneous servers}} & \multicolumn{2}{c|}{\textbf{Heterogeneous servers}} \\
		\cline{3-6}
		&\textbf{distribution}  & \textbf{i.i.d. copies} &\textbf{ identical copies} &\textbf{ i.i.d. copies} &\textbf{ identical copies}\\
		\hline
		\multirow{3}{*}{\textbf{FCFS}} &\textbf{Exponential}  & General red., \cite{Gardner16} & Redundancy-$d$, \cite{Anton2019} & General red.,\cite{Gardner16} &  \\
		\cline{2-6}
		& \multirow{2}{*}{\textbf{Scaled Bernoulli}} & Redundancy-$d$, \cite{Raaijmakers2018} &  & &  \\
		&  & (Asymptotic regime) &  &  &  \\
		\cline{1-6}
		\multirow{3}{*}{\textbf{PS}} & \textbf{Exponential}  & Redundancy-$d$, \cite{Anton2019} & Redundancy-$d$, \cite{Anton2019}  &  &  \\
		\cline{2-5}\Cline{6-6}{1pt}
		& 	\multirow{2}{*} {\textbf{General}}  &  Redundancy-$d$, \cite{Raaijmakers2019} &  Redundancy-$d$, \cite{Raaijmakers2019}  &  & 
		\multicolumn{1}{!{\vrule width 1pt}c!{\vrule width 1pt}}{General red.} \\
		& & (Necessary condition) & & & \multicolumn{1}{!{\vrule width 1pt}c!{\vrule width 1pt}}{(Light-tailed)}\\
		\cline{1-5}\Cline{6-6}{1pt}
		\multirow{1}{*}{\textbf{ROS}} & \textbf{Exponential}  & Redundancy-$d$, \cite{Anton2019} & Redundancy-$d$, \cite{Anton2019} &   &  \\
		\hline		
	\end{tabular}
\end{table*}

\section{Model description}
\label{sec:model}
We consider a $K$ parallel-server system with heterogeneous capacities $\mu_k$, for $k=1,\ldots,K$. Each server has its own queue, where Processor Sharing (PS) service policy is implemented. 
We denote by $S=\{1,\ldots,K\}$ the set of all servers.

\begin{figure*}
	\centering
	\includegraphics[scale=0.8]{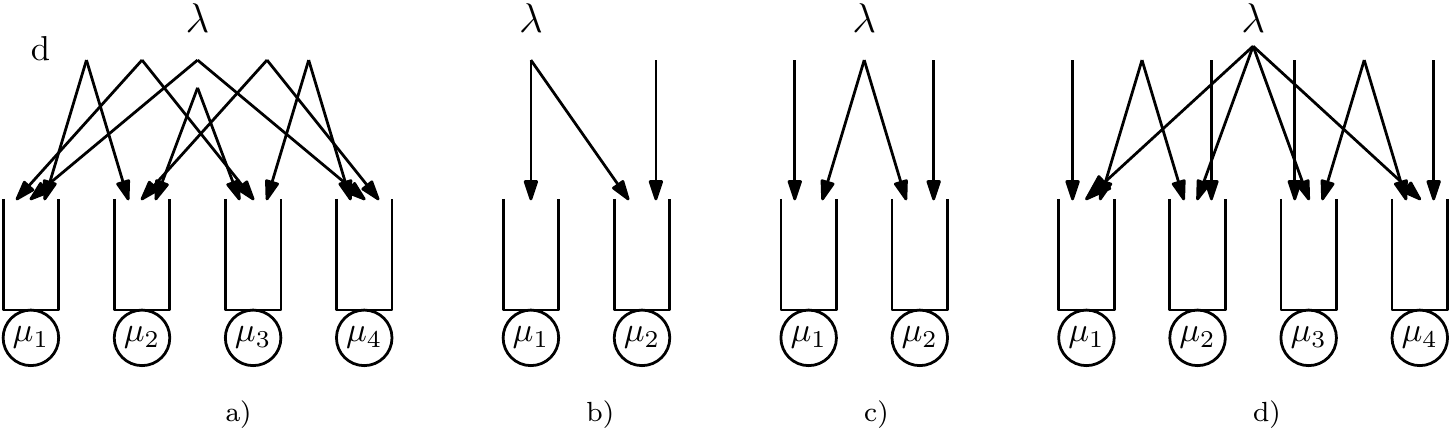}
	\caption{  From left to right, the redundancy-$d$ model (for $K=4$ and $d=2$), the $N$-model, the $W$-model and the $WW$-model.}
	\label{fig:red_examples}
\end{figure*}

Jobs arrive to the system according to a Poison process of rate~$\lambda$. 
Each job is labelled with a type $c$ that represents the subset of compatible servers to which type-$c$ jobs can be sent: i.e., $c=\{s_1,\ldots s_n\}$, where $n\leq K$, $s_1,\ldots,s_n\in S$ and $s_i\neq s_l$, for all $i\neq l$. A job is with probability $p_c$ of type~$c$, where $\sum_{c\in\mathcal C} p_c=1$. 
We denote by $\mathcal C$ the set of all types in the system, i.e, $\mathcal C=\{ c\in \mathcal P(S) \ : \ p_c>0\}$, where $\mathcal P(S)$ contains all the possible subsets of~$S$.
Furthermore, we denote by $\mathcal C(s)$ the subset of types that have server  $s$ as compatible server, that is, $\mathcal C(s)=\{c\in \mathcal C : s\in c\}$. For instance, the $N$-model is a two-server system with jobs of types $c=\{2\}$ and $c=\{1,2\}$, see Figure~\ref{fig:red_examples} b). Thus, $\mathcal C=\{\{2\},\{1,2\}\}$, $\mathcal C(1)=\{\{1,2\}\}$ and $\mathcal C(2)=\{\{2\},\{1,2\}\}$, with $p_{\{2\}},p_{\{1,2\}}>0$.

Job sizes are distributed according to a general random variable $X$ with cumulative distribution function $F$ and unit mean.
Additionally, we assume that
\begin{enumerate}
	\item $F$ has no atoms.
	\item $F$ is a light tailed distribution in the following sense,
	\begin{equation}\label{eq:lt:1}
	\lim\limits_{r\to\infty} \sup_{a\geq 0} \mathbf E[(X-a)1_{\{X-a>r\}}\vert X>a]=0.
	\end{equation}
\end{enumerate}
\begin{remark}
	These technical conditions have been used previously in the literature to prove stochastic stability from fluid limits arguments (see \cite{Lee2008} and \cite{PFTA12}) in the context of processor sharing networks and cannot be avoided easily.
	However, it can be seen (as observed in \cite{PFTA12}) that Equation \eqref{eq:lt:1} also implies
	\begin{equation}\label{eq:lt:2}
	\sup_{a\geq 0}  \mathbf E[(X-a)\vert X>a]\leq\Phi<\infty,
	\end{equation} 
	which is a usual light tail condition (see \cite{Foss2013}).
	Hence, Equations \eqref{eq:lt:1} and \eqref{eq:lt:2} though exclude heavy tail distributions like Pareto, include large sets of distributions 
	as phase type (which are dense in the set of all distributions on $\mathbb R^+$), distributions with bounded support, exponential and hyper-exponential distributions.

\end{remark}
We consider two load balancing policies, which determine how the jobs are dispatched to the servers. Note that both load balancers are oblivious to the capacities of the servers. 
\begin{itemize}
	\item Bernoulli routing: a type-$c$ job is send with uniform probability to one of its compatible servers in $c$. 
	\item Redundancy model: a type-$c$ job sends  identical copies to its  $|c|$ compatible servers. That is, all the copies of a job have exactly the same size.
	The job (and  corresponding copies) departs the system when one of its copies completes service. 
\end{itemize} 


In this paper, we will study the stability condition under both load balancing policies. We call the system stable when the underlying process is positive Harris recurrent, and unstable when the process is transient. A stochastic process is positive Harris recurrent if there exists a petite-set $C$ for which $P(\tau_C<\infty)=1$ where $\tau_C$ is the stopping time of $C$, see e.g.,  \cite{Bramson06,Asmussen2002,Meyn1993} for the corresponding definitions. We note that when the state descriptor is Markovian, positive Harris recurrent is equivalent to positive recurrent.


We define $\lambda^{R}$ as the  value of $\lambda$ such that the redundancy model is stable if $\lambda < \lambda^{R}$ and unstable if  $\lambda > \lambda^{R}$. Similarly, we define $\lambda^{B}$ for the Bernoulli routing system. We aim to  characterize when  $\lambda^{R} > \lambda^{B}$, that is, when does redundancy improve the stability condition compared to no redundancy.

For Bernoulli,  $\lambda^{B}$ can be easily found. Under Bernoulli routing, a job chooses a server uniformly at random, hence, type-$c$ jobs arrive at server $s$ at rate $\lambda p_c/\vert c\vert$. Thus, the Bernoulli system reduces to  $K$ independent servers, where server~$s$ receives arrivals at rate $\lambda(\sum_{c\in\mathcal C(s)} \frac{p_c}{\vert c\vert})$ and has a departure rate $\mu_s$, for all $s\in S$. The stability condition is hence, 
\begin{equation}\label{stab:g-non}
\lambda < 
\lambda^{B}=\min_{s\in S} \left\{\frac{\mu_s}{\sum_{c\in\mathcal C(s)} \frac{p_c}{\vert c \vert}}\right\}.\end{equation}

In order to characterize $\lambda^{R}$, we need to study the system under redundancy in more detail. For that, we denote by $N_c(t)$  the number of type-$c$ distinct jobs that are present in the redundancy system at time $t$ and $\vec N(t)=(N_c(t),c\in \mathcal{C})$. Furthermore, we denote the number of copies per server by $M_s(t) := \sum_{c\in \mathcal{C}(s)} N_c(t)$, $s\in S$, and $\vec M(t)=(M_1(t),\ldots,M_K(t))$. For the $j$-th type-$c$ job, let $b_{cj}$ denote the service requirement of this job, for $j=1,\ldots,N_c(t)$, $c\in\mathcal C$.
Let $a_{cjs}(t)$ denote the attained service in server $s$ of the $j$-th type-$c$ job at time $t$. We denote by $ A_c(t)=(a_{cjs}(t))_{js}$ a matrix on $\mathbb{R}_+$ of dimension $N_c(t)\times \vert c\vert$. Note that the number of type-$c$ jobs increases by one at rate $\lambda p_c$, which implies that a row composed of zeros is added to $A_c(t)$. When one element $a_{cjs}(t)$ in matrix $A_c(t)$ reaches the required service $b_{cj}$, the corresponding job departs and all of its copies are removed from the system. Hence, row~$j$ in matrix $A_c(t)$ is removed. 
We further let $\phi_s(\vec M(t))$ be the capacity that each of the copies in server~$s$ obtains when in state $\vec M(t)$, which under PS is given by, 
$\phi_s(\vec M(t)):=\frac{\mu_s}{M_s(t)}$. 
The cumulative service that a copy in server $s$ gets during the time interval $(v,t)$ is 
$$ \eta_s(v,t):=\int_{x=v}^t \phi_s(\vec M(x)) \textrm dx.$$

In order to characterize the stability condition, we define the  \emph{capacity-to-fraction-of-arrivals ratio} of a server in a subsystem: 
\begin{definition}[Capacity-to-fraction-of-arrival ratio]\label{def:cpr}
	For any given set of servers $\tilde S\subseteq S$ and its associated set of job types $\tilde{\mathcal C}=\{c\in\mathcal C \ : \ c\subseteq \tilde S\}$, the \emph{capacity-to-fraction-of-arrival ratio} of server $s\in\tilde S$ in this so-called  $\tilde S$-subsystem is defined by 
	$
	\frac{\mu_s}{\sum_{c\in\tilde{\mathcal C}(s)} p_c},
	$
	where $\tilde{\mathcal C}(s)=\tilde{\mathcal C}\cap \mathcal C(s)$ is the subset of types in $\tilde{\mathcal C}$ that are served in server~$s$.
\end{definition}



\subsection*{Some common models}
A well-known   structure is  the \emph{redundancy-$d$}  model, see Figure~\ref{fig:red_examples} a). Within this model, each job has  $d$ out of $K$ compatible servers, where $d$ is fixed. That is, $p_c>0$ for all $c\in \mathcal P(S)$ with $\vert c\vert=d$, and $p_c=0$ otherwise,  
so that there are  
$\vert\mathcal C\vert=\binom{K}{d}$ types of jobs. 
If additionally, $p_c=1/\binom{K}{d}$ for all $c\in\mathcal C$, we say that the arrival process of jobs is homogeneously distributed over types. We will call this model the \emph{redundancy}-$d$ model with homogeneous arrivals. The particular case where server capacities are also homogeneous, i.e.,  $\mu_k=\mu$ for all $k=1,\ldots,K,$
will be called the \emph{redundancy}-$d$ model with homogeneous arrivals and server capacities. 

In~\cite{Gardner2018} the  \emph{nested} redundancy model was introduced, where for all $c,c'\in\mathcal C$, either \emph{i)} $c\subset c'$ or \emph{ii)} $c'\subset c$ or \emph{iii)} $c\cap c'=\emptyset$. First of all, note  that the redundancy-$d$ model does not fit in the nested structure.   The smallest nested system is the so called $N$-model (Figure~\ref{fig:red_examples} b)):  this is a $K=2$ server system with types $\mathcal C=\{\{2\},\{1,2\}\}$.   Another nested system is the $W$-model (Figure~\ref{fig:red_examples} c)), that is, $K=2$ servers and types $\mathcal C = \{\{1\}, \{2\}, \{1,2\}\}$. 
In  Figure~\ref{fig:red_examples} d), a nested model with $K=4$ servers and $7$ different jobs types, $\mathcal C=\{\{1\},\{2\},\{3\},\{4\},$ $\{1,2\},\{3,4\},\{1,2,3,4\}\}$ is given. This model is referred to as  the $WW$-model. 

\section{An illustrative example}
\label{sec:example}

\begin{figure}[h]
	\begin{subfigure}{.495\textwidth}
		\centering
		\includegraphics[width=.8\textwidth]{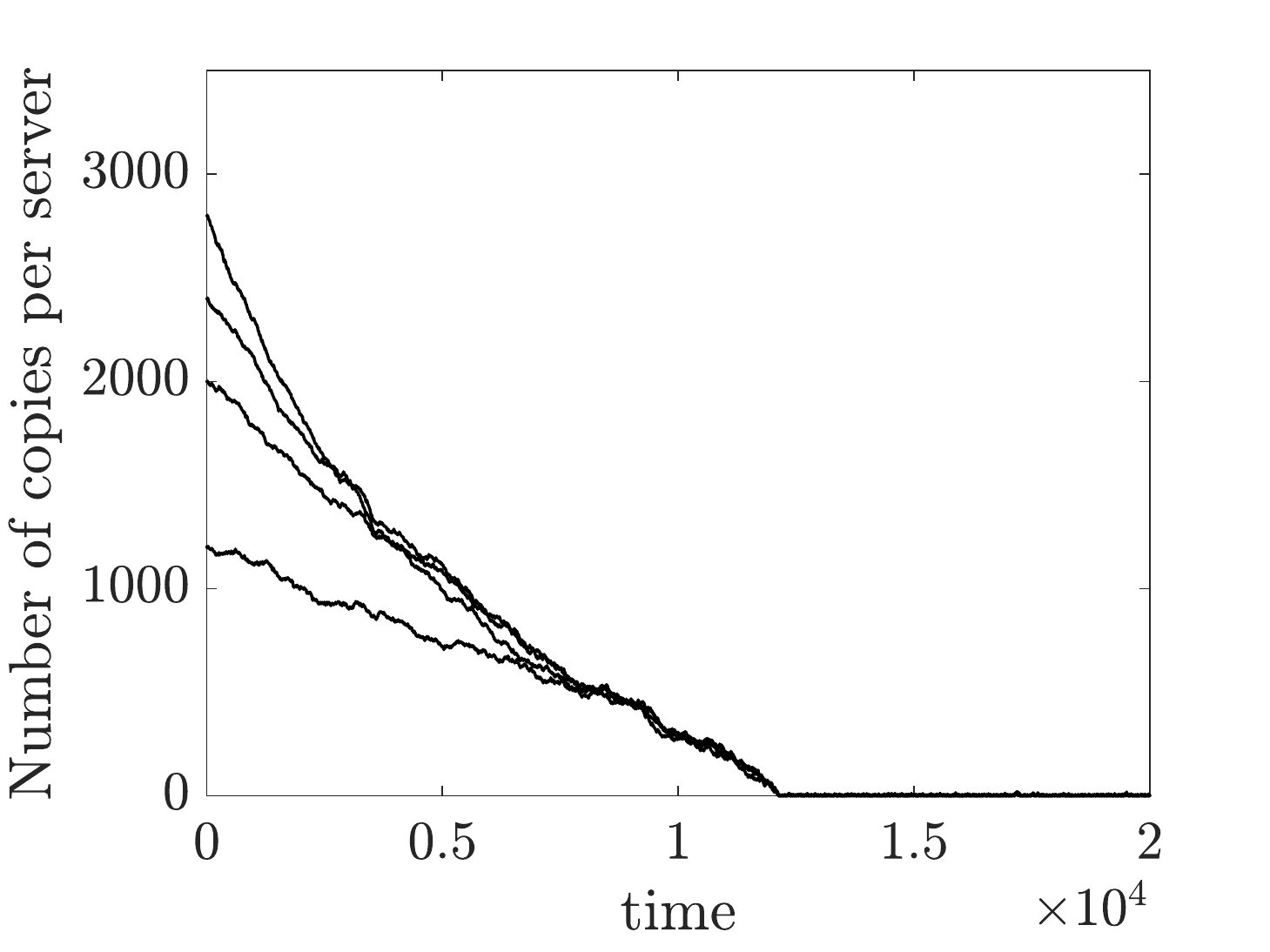}
		
		a) $\lambda = 1.8$
	\end{subfigure}
	\begin{subfigure}{.495\textwidth}
		\centering
		\includegraphics[width=.8\textwidth]{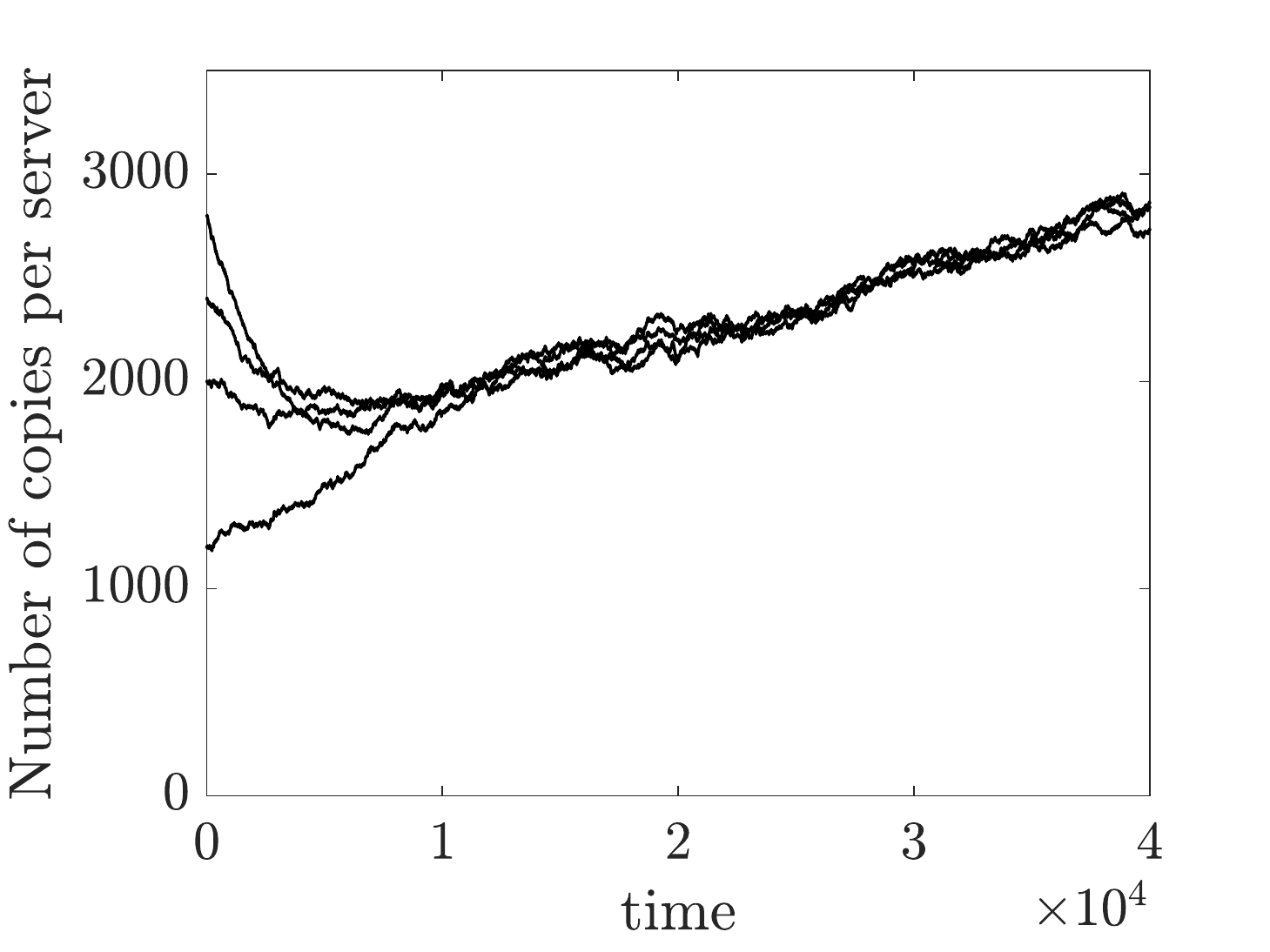}
		
		b) $\lambda = 2.1$
	\end{subfigure}
	
	\begin{subfigure}{.495\textwidth}
		\centering
		\includegraphics[width=.8\textwidth]{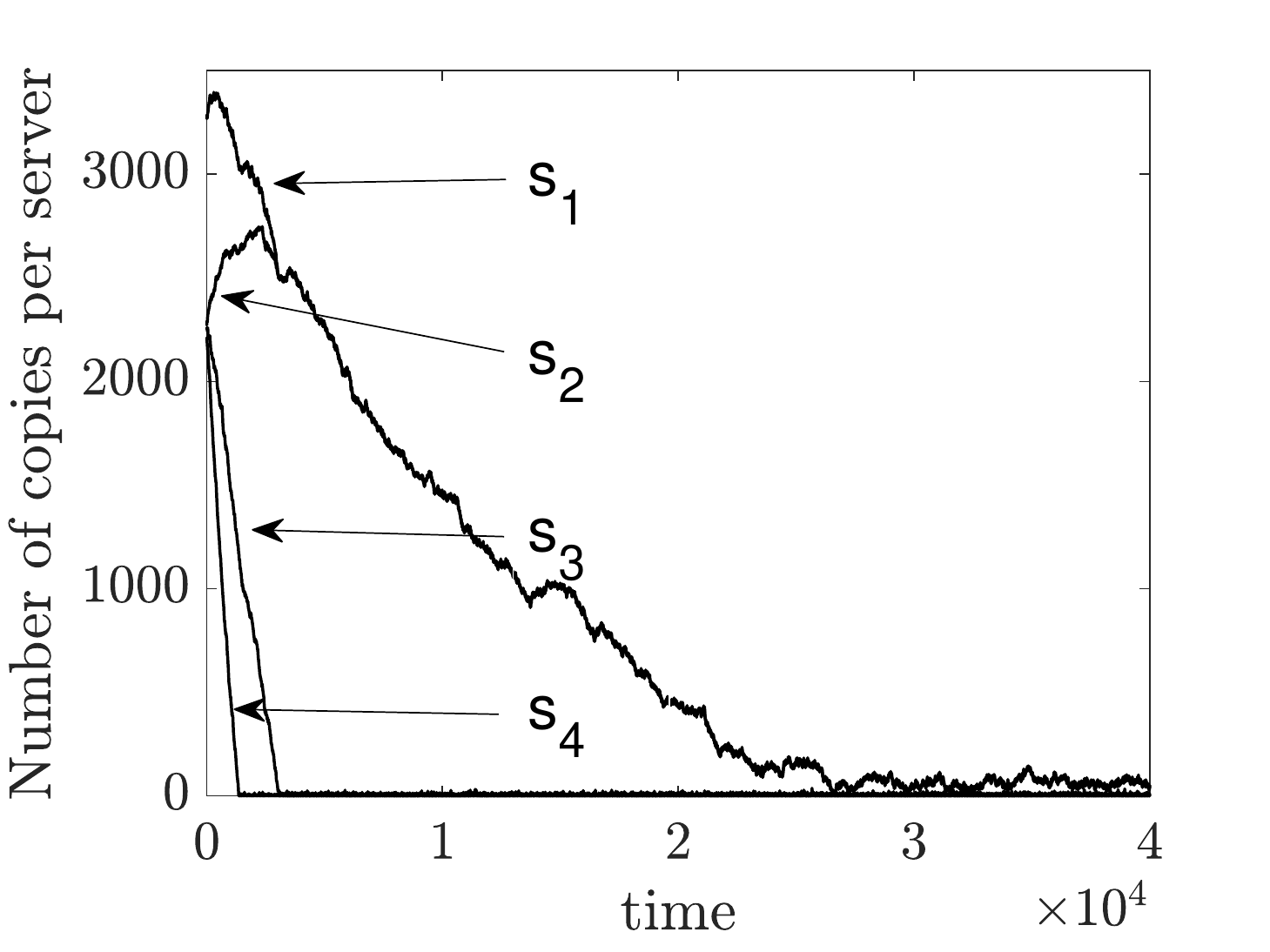}
		
		c) $\lambda =7.5$ 	
	\end{subfigure}
	\begin{subfigure}{.495\textwidth}
		\centering
		\includegraphics[width=.8\textwidth]{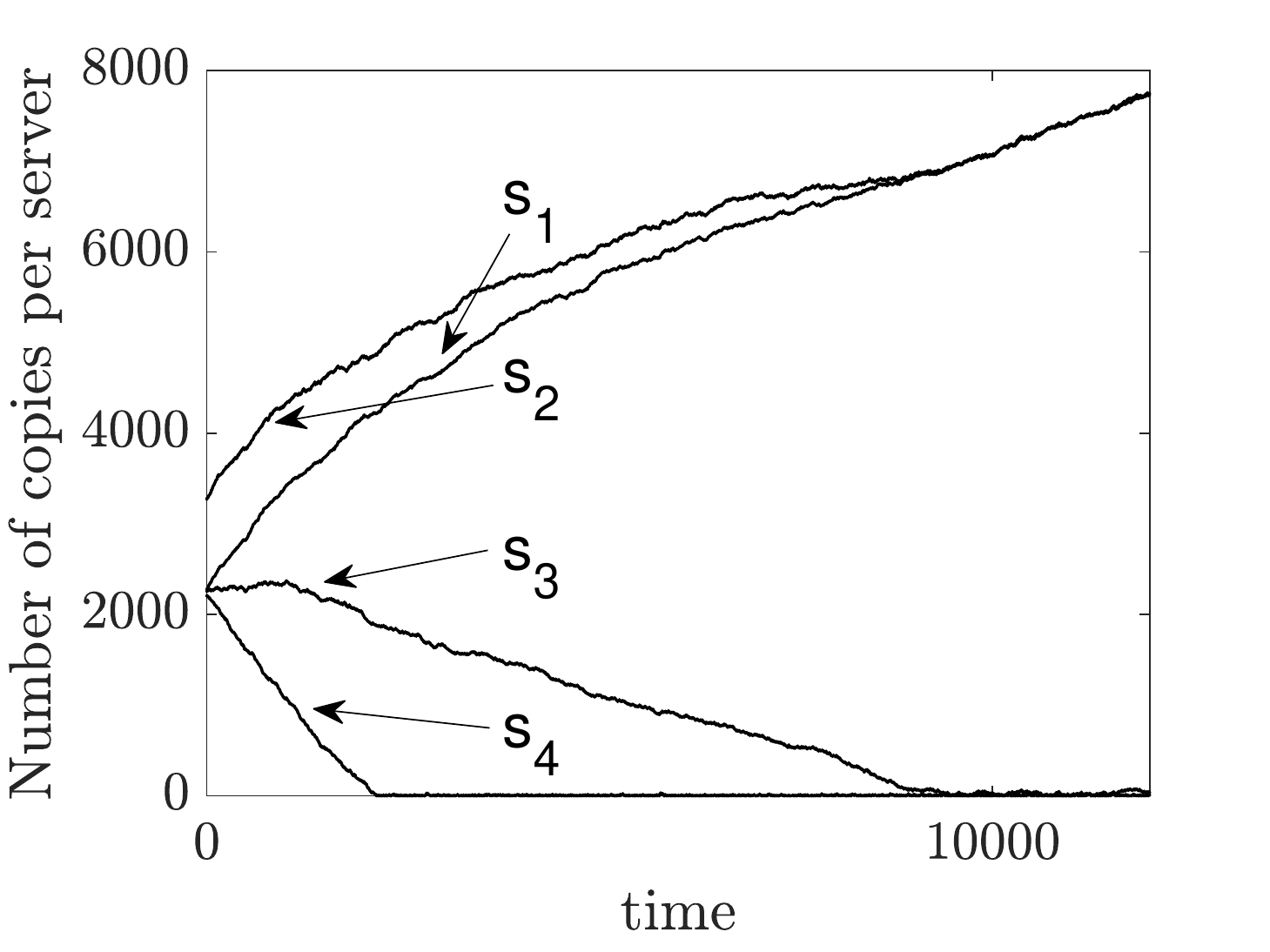}
		
		d) $\lambda = 9$
	\end{subfigure}
	\caption{Trajectory of the number of copies per server with respect to time for a $K=4$ redundancy-$2$ system with exponentially distributed job sizes.
		Figures a) and b) consider homogeneous capacities $\mu_k=1$ for $k=1,\ldots,4$ and homogeneous arrival rates per type, $p_c=1/6$ for all $c\in\mathcal C$, with a) $\lambda=1.8$ and b) $\lambda=2.1$. Figures c) and d) consider heterogeneous server capacities $\vec \mu=(1,2,4,5)$ and arrival rates per type $\vec p=(0.25, 0.1, 0.1, 0.2, 0.2, 0.15)$  for types $\mathcal C$, c) with $\lambda=7.5$ and d) $\lambda=9$.}
	\label{Fig:event}
\end{figure}

Before formally stating the main results in Section~\ref{subsec:stab}, we first illustrate through a numerical example some of the key aspects of our proof, and in particular the essential role played by the capacity-to-fraction-of-arrival ratio defined in Definition~\ref{def:cpr}. In Figure~\ref{Fig:event} we plot the trajectories of the number of copies per server with respect to time for a $K=4$ redundancy-$2$ system (Figure~\ref{fig:red_examples} a)), that is 
$\mathcal C=\{ \{1,2\},\{1,3\},\{1,4\},\{2,3\},\{2,4\},\{3,4\} \}$. Our proof techniques will rely on fluid limits, and therefore we chose large initial points. 
Figures~\ref{Fig:event} a) and b) show the trajectories  when servers and arrivals of types are homogeneous
for $\lambda=1.8$   and $\lambda = 2.1$, respectively.
Figures~\ref{Fig:event} c) and d) consider a heterogeneous system (parameters see the legend) 
for  $\lambda=7.5$  and $\lambda=9$, respectively.


The homogeneous example (Figure~\ref{Fig:event} a) and b)) falls  within the scope  of~\cite{Anton2019}. 
There it is shown that the stability condition is $\lambda < \frac{\mu K}{d}$. 
We note that this condition coincides with  the stability condition of a system in which all the $d$ copies need to be fully served. 
In Figure~\ref{Fig:event} a) and b), the value for $\lambda$ is chosen such that they represent a stable and an unstable system, respectively.
As formally proved in \cite{Anton2019}, at the fluid scale, when the system is stable the largest queue length   decreases, whereas in the unstable case the minimum queue length increases. It thus follows, that in the homogeneous case, either \emph{all classes} are stable, or unstable.  

The behavior of the heterogeneous case is rather different. The parameters corresponding to Figures~\ref{Fig:event} c) and d) are such that the system is stable in c), but not in d). In Figure~\ref{Fig:event} c) we see that the trajectories of all queue lengths are not always decreasing, including the maximum queue length. In Figure~\ref{Fig:event} d), we observe that the number of copies in  servers 3 and 4 are decreasing, whereas those of servers 1 and 2 are increasing. 

When studying stability for the heterogeneous setting, one needs to reason recursively.
First, assume that each server $s$ needs to handle its full load, i.e., $\lambda \frac{\sum_{c\in \mathcal{C}(s) p_c}}{\mu_s}$. Hence, one can simply compare the servers capacity-to-fraction-of-arrival ratios, $\mu_s/\sum_{c\in \mathcal{C}(s)} p_c$, to see which server is the least-loaded server and could hence potentially empty first. In this example, server 4 has the maximum capacity-to-fraction-of-arrival ratio, and, in fluid scale, will reach zero in finite time, and remain zero, since $\mu_4/\sum_{c\in \mathcal{C}(4)} p_c = 
5/(p_{\{1,4\}}+p_{\{2,4\}}+p_{\{3,4\}})=11.11$ is larger than  $\lambda=7.5$. 


Whenever, at fluid scale, server 4 is still positive,  the other servers might either increase or decrease. 
However, the key insight is that once the queue length of server 4 reaches 0, the fluid behavior of the other classes no longer depend on the jobs that also have server~4 as compatible server. That is, we are sure that all jobs that have server~4 as compatible server, will be fully served in server~4, since server~4 is in fluid scale empty and all the other servers are overloaded. Therefore, jobs with server 4 as compatible server can be ignored, and we are left 
with a subsystem formed by servers $\{1,2,3\}$ and without the job types served by server 4.  
Now again, we consider the maximum capacity-to-fraction-of-arrival ratio in order to determine the least-loaded server, but now for the subsystem $\{1,2,3\}$. This time, server~3 has the maximum capacity-to-fraction-of-arrival ratio, which is $4/(p_{\{1,3\}}+p_{\{2,3\}})=10$. Since this value is larger than $\lambda=7.5$, it is a sufficient condition for server 3 to empty. 

Similarly, once server 3 is empty, we consider the subsystem with servers 1 and 2 only. Hence, there is only one type of jobs, $\{1,2\}$.  Now server~2 is the least-loaded server and its capacity-to-fraction-of-arrival ratio is $2/p_{\{1,2\}}=8$. This value being larger than the arrival rate, implies that   server 2 (and hence server 1, because there is only one job type) will be stable too. 
Indeed, in Figures~\ref{Fig:event} c) we also observe that as soon as the number of copies in server~3 is relatively small compared to that of server 1 and server 2, the number of copies in both server 1 and server 2 decreases.


We can now explain the evolution observed in Figure~\ref{Fig:event} d) when $\lambda=9$. The evolution for servers 4 and 3 can be argued as before: both their capacity-to-fraction-of-arrival ratios are larger than $\lambda = 9$, hence they empty in finite time. However, the capacity-to-fraction-of-arrival ratio of the subsystem with servers 1 and 2, which is 8, is strictly smaller than the arrival rate.  We thus observe that, unlike in the homogeneous case, in the heterogeneous case some servers might be stable, while others (here server 1 and 2) are unstable. 

Proposition~\ref{rem:stab} formalizes the above intuitive explanation, by showing that the stability of the system can be derived recursively.

The capacity-to-fraction-of-arrival ratio allows us now to reinterpret the homogeneous case depicted in Figure~\ref{Fig:event} a) and b). In this case,  the capacity-to-fraction-of-arrival ratio of all the servers is the same, which implies \emph{(i)} that either all servers will be stable, or all unstable, and \emph{(ii)} from the stability viewpoint is as if all copies received service until completion.

\section{Stability condition}
\label{sec:gen_red}

\subsection{Multi-type job multi-type server system}
\label{subsec:stab}

In this section we discuss the stability condition of the general redundancy system with PS. 
In order to do so, we first define several sets of  subsystems, similar to as what we did in the illustrative example of Section~\ref{sec:example}. 

The first subsystem includes all servers, that is $S_1=S$. 
We denote by $\mathcal L_1$ the set of servers with highest capacity-to-fraction-of-arrival ratio in the system~$S_1=S$. Thus, 
$$\mathcal L_1=\left\{s\in S_1\ : \ s=\arg\max_{\tilde s\in  S_1} \left\{\frac{\mu_{\tilde s}}{\sum_{c\in{\mathcal C}} p_c} \right\}\right\}.$$ 

For $i=2,\ldots, K$, we define recursively 
\begin{eqnarray*}
	S_i&:=& S\backslash \cup_{l=1}^{i-1}\mathcal L_l,\\
	{\mathcal C}_i&:=&\{c\in\mathcal C \ : \ c\subset S_i\},\\
	\mathcal C_i(s)&:=&\mathcal C_i\cap\mathcal C(s),\\
	\mathcal L_i&:=&\left\{s\in S_i \ : \ s=\arg\max_{\tilde s\in  S_{i}} \left\{\frac{\mu_{\tilde s}}{\sum_{c\in{\mathcal C}_{i}(\tilde s)} p_c} \right\} \right\}.
\end{eqnarray*}
The   $S_i$-subsystem will refer to the system consisting of the servers in   $S_i$, with only jobs of types in the set ${\mathcal C}_i$. 
The $\mathcal C_i(s)$ is the subset of types that are served in server $s$ in the $S_i$-subsystem. We let $ \mathcal C_1=\mathcal C$. The  $\mathcal L_i$ represents the set of servers $s$ with highest capacity-to-fraction-of-arrival ratio in the $S_i$-subsystem, or in other words, the least-loaded servers in the $S_i$-subsystem.
Finally, we denote by $i^*:= \arg\max_{i=1,\ldots, K}\{\mathcal C_i \ : \ \mathcal C_i\neq\emptyset\}$ the last index $i$ for which the  subsystem $S_i$ is not empty of job types.

\begin{figure}[t]
	\centering
	\includegraphics[width=0.8\textwidth]{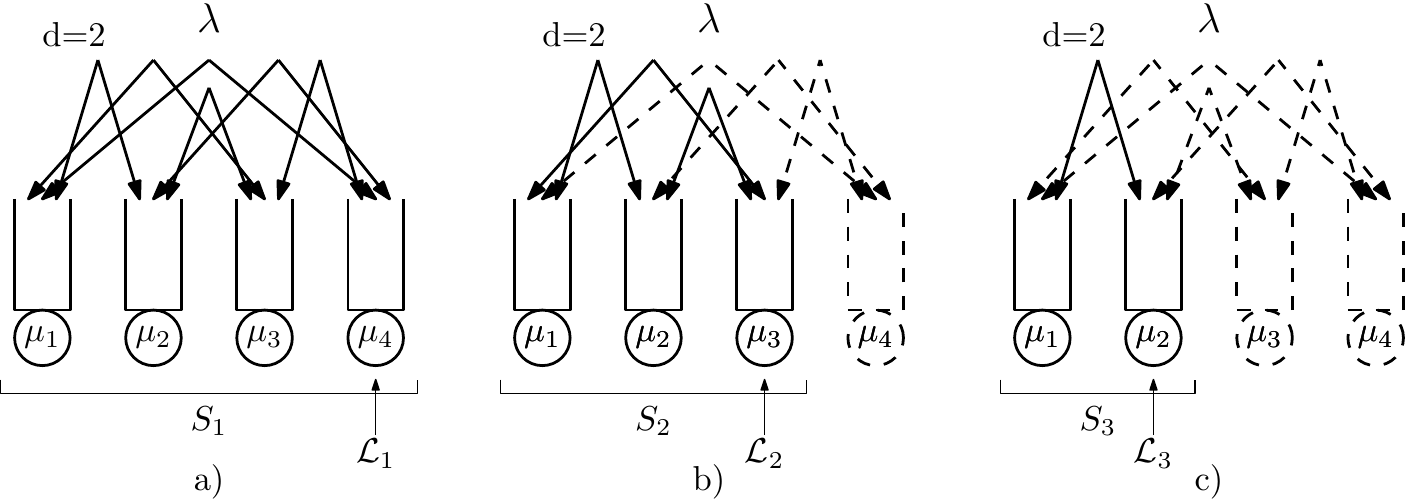}
	\caption{$K=4$ server system under redundancy-$2$. In a) subsystem $S_1$, in b) subsystem $S_2$ and in c) subsystem $S_3$.}
	\label{Fig_red_2}
\end{figure}

\begin{remark}
	We illustrate the above definitions by applying them to the particular example considered in Section~\ref{sec:example}.
	The first subsystem consists of servers $S_1=S=\{1,2,3,4\}$ and all job types, see Figure~\ref{Fig_red_2} a). The capacity-to-fraction-of-arrival ratios in the $S_1$ subsystem are: $\{2.2,3.07,8.8,11.1\}$, and thus $\mathcal{L}_1=\{4\}$. The second subsystem is formed by  $S_2=\{1,2,3\}$ and job types that are compatible with  server 4 can be ignored, that is, $\mathcal C_2=\{\{1,2\}, \{1,3\}, \{2,3\}\}$, see Figure~\ref{Fig_red_2} b). The capacity-to-fraction-of-arrival ratios for servers in the $S_2$ subsystem are given by $\{2.8,4.4,10\}$, and thus $\mathcal L_2=\{3\}$. The third subsystem consists  of servers $S_3=\{1,2\}$ and job types that are compatible with servers 3 or 4 can be ignored, that is, $\mathcal C_3=\{\{1,2\}\}$, see Figure~\ref{Fig_red_2} c). The capacity-to-fraction-of-arrival ratios for servers in the $S_3$ subsystem are given by $\{4,8\}$. Hence, $\mathcal L_3=\{2\}$. Then, $S_4=\{1\}$, but $\mathcal C_4=\emptyset$, so that $i^*=3$.
\end{remark}

The value of the highest capacity-to-fraction-of-arrival ratio in the $S_i$-subsystem is denoted by
\begin{equation*}
	CAR_i:=\max_{\tilde s\in  S_{i}} \{\frac{\mu_{\tilde s}}{\sum_{c\in{\mathcal C}_{i}(\tilde s)} p_c}\}, \ \mbox{for } i=1,\ldots, i^*.
\end{equation*}
Note that $CAR_i  = \frac{\mu_{s}}{\sum_{c\in{\mathcal C}_{i}(s)} p_c},$  for any   $s\in \mathcal{L}_i.$ 

In the following proposition we characterize the stability condition for servers in terms of the capacity-to-fraction-of-arrival ratio corresponding to each subsystem. It states that servers that have highest capacity-to-fraction-of-arrival ratio in subsystem~$S_i$ can  be stable if and only if all servers in $S_1,\ldots, S_{i-1}$ are stable as well. The proof can be found in Section~\ref{sec:stab_cond_proof}.

\begin{proposition}\label{rem:stab}
	For a given $i\leq i^*$, servers $s\in\mathcal L_{i}$ are stable if  $\lambda < CAR_l$, for all $l=1,\ldots,i$.  Servers $s\in\mathcal L_{i}$ are unstable if there is an $l=1,\ldots,i$ such that $\lambda > CAR_l$.
\end{proposition}


\begin{corollary}\label{stab:cond}
	The redundancy system is stable if
	$\lambda<CAR_i$,   for all $i=1,\ldots, i^*.$
	The  redundancy system is unstable if there exists an $\iota\in\{1,\ldots, i^*\}$ such that $\lambda > CAR_\iota$.
\end{corollary}

We note that  $CAR_l$, $l=1,\ldots,i$, are not necessarily ordered with respect to $l$.
From the corollary, we hence obtain that the stability region under redundancy is given by
\begin{equation}
\label{eq:lR}
\lambda^{R}= \min_{i=1,\ldots, i^*}CAR_i.
\end{equation}

We now write an equivalent representation of the stability condition (proof see Appendix).  Denote by $\mathcal R(c)$ the set of  servers  where  type-$c$ jobs achieve maximum capacity-to-fraction-of-arrival ratio, or in other words, the set of least-loaded servers for type $c$:
$$\mathcal  R(c):=\{s: \exists i, \mbox{ s.t. } c\in C_i(s) \mbox{ and } s\in \mathcal{L}_i\}.$$
Note that there is a unique  subsystem $S_i$ for which this happens, i.e., $\mathcal  R(c)\subseteq \mathcal{L}_i$ for exactly one $i$. We note that for a type-$c$ job, if $c$ contains at least a server that was removed in the $i$th iteration, then $\mathcal R(c)\subseteq \mathcal L_i$. We further let $\mathcal{R}:=\cup_{c\in \mathcal{C}}\mathcal R(c)$.


\begin{corollary}\label{stab:cond2}
	The redundancy system is stable if
	$\lambda \sum_{c: s\in \mathcal{R}(c)} p_c <\mu_s$, for all $ s\in \mathcal R$.
	The redundancy system is unstable if there exists an $s\in \mathcal R$ such that $\lambda \sum_{c: s\in \mathcal R(c)} p_c >\mu_s$.
\end{corollary}

From the above corollary, we directly observe that the stability condition for the redundancy system coincides with the   stability condition corresponding to $K$~individual servers where each  type-$c$ job is only dispatched to its least-loaded servers.

\subsection{Particular redundancy structures}
In this subsection we discuss the stability condition for some particular cases of redundancy: redundancy-$d$ and nested systems. 

\subsection*{Redundancy-$d$}
\label{subsec:red_d}
We focus here on the redundancy-$d$ structure (defined in Section~\ref{sec:model}) with homogeneous arrivals, i.e. $p_c=\frac{1}{\binom{K}{d}}$ for all $c\in\mathcal C$.

In case the servers capacities are homogeneous, $\mu_k=\mu$ for all $k$, the model fits in the setting of~\cite{Anton2019} where it was proved to be stable if $\lambda d<\mu K$.  This would also follow from Corollary~\ref{stab:cond}: 
Since arrivals are homogeneous,    the arrival rate to each server is $\lambda d/K$, thus the capacity-to-fraction-of-arrival ratio at every server is $\mu K/d$. 
This implies that $\mathcal L_1=S$, $i^*=1$ and $\mathcal R(c)=c$ for all $c\in\mathcal C$. From Corollary~\ref{stab:cond}, we obtain that the system is stable if $\lambda d<\mu K$. 

For heterogeneous servers capacities, which was not studied in~\cite{Anton2019}, we have the following:
\begin{corollary}\label{corol:homo}
	Under redundancy-$d$ with homogeneous arrivals and  $\mu_1<\ldots<\mu_K$, the system is stable if for all  $i=d,\ldots,K$, $\lambda \frac{\binom{i-1}{d-1}}{\binom{K}{d}} < \mu_{i}$. The system is unstable if there exists $i\in\{d,\ldots,K\}$ such that $\lambda \frac{\binom{i-1}{d-1}}{\binom{K}{d}} > \mu_{i}$.
\end{corollary}

In the homogeneous case, it is easy to deduce that the stability condition,  $\lambda d<\mu K$, decreases as $d$ increases. However, in the heterogeneous case, both the numerator and denominator are non-monotone functions of $d$, and as a consequence it is not straightforward how the stability condition depends on $d$. This dependence on~$d$ will be numerically studied in Section~\ref{sec:impr_d}.

\subsection*{Nested systems}
\label{subsec:nested}
In this section we consider two nested redundancy systems.  

\subsubsection{$N$-model}
\label{sec:N}
The simplest nested model is the $N$-model. This is a $K=2$ server system with capacities $\vec\mu=\{\mu_1,\mu_2\}$ and types $\mathcal C=\{\{2\},\{1,2\}\}$, see Figure~\ref{fig:red_examples} (b). A job is of type $\{2\}$ with probability $p$ and of type $\{1,2\}$ with probability $1-p$. The stability condition is $\lambda<\lambda^{R}$ where: 
$$ \lambda^{R}=\left\{\begin{array}{ll} \mu_2,& 0\leq p\leq \frac{\mu_2-\mu_1}{\mu_2} \\ 
\mu_1/(1-p), & \left(\frac{\mu_2-\mu_1}{\mu_2}\right)^+ \leq p\leq\frac{\mu_2}{\mu_1+\mu_2}\\
\mu_2/p, & \frac{\mu_2}{\mu_1+\mu_2}<p\leq1.
\end{array}\right.$$
The above is obtained as follows: The capacity-to-fraction-of-arrival ratio of the system is $\mu_1/(1-p)$ and $\mu_2$,  respectively for server~1 and server~2. First assume  $\mu_1/(1-p)>\mu_2$. Then $\mathcal L_1=\{1\}$ and the second \ subsystem is composed of server $S_2=\{2\}$ and $\mathcal C_2=\{\{2\}\}$, with arrival rate $\lambda p$ to server 2. Hence the capacity-to-fraction-of-arrival ratio of server 2 in the $S_2$-subsystem is $\mu_2/p$. From Corollary~\ref{stab:cond}, it follows that $\lambda^{R}=\min\{\mu_1/(1-p),\mu_2/p\}$. 
On the other hand, if $\mu_1/(1-p)<\mu_2$, then $\mathcal L_1=\{2\}$, and $S_2=\{1\}$, but $\mathcal C_2=\emptyset$. Thus, $\lambda^{R}=\mu_2$. Lastly, if $\mu_1/(1-p)=\mu_2$, $\mathcal L_1=\{1,2\}$, thus $S_2=\emptyset$ and $\mathcal C_2=\emptyset$. Hence, $\lambda^{R}=\mu_2$.

We observe that the stability condition $\lambda^{R}$, is a continuous function reaching the maximum value $\lambda^{R}=\mu_1+\mu_2$ at $p=\mu_2/(\mu_1+\mu_2)$.  
It thus follows that for $p=\mu_2/(\mu_1+\mu_2)$, redundancy achieves the maximum stability condition. We note however that in this paper our focus is not on finding the best redundancy probabilities, but instead whether given the  probabilities $p_c$ --which are determined by the characteristics of the job types and matchings -- the system can benefit from redundancy.

\subsubsection{$W$-model}
The $W$-model is a $K=2$ server system with capacities $\vec\mu=\{\mu_1,\mu_2\}$ and types $\mathcal C=\{\{1\},\{2\},\{1,2\}\}$, see Figure~\ref{fig:red_examples} c). A job is of type $\{1\}$ with probability $p_{\{1\}}$, type $\{2\}$ with probability $p_{\{2\}}$ and of type $\{1,2\}$ with probability $p_{\{1,2\}}$. W.l.o.g., assume $(1-p_{\{2\}})/\mu_1\geq (1-p_{\{1\}})/\mu_2$, that is,  the load on server~1 is larger than or equal to that on server~2.  The stability condition is then given by:
$$ \lambda^{R}=\left\{\begin{array}{ll}
\mu_2/(1-p_{\{1\}}),& p_{\{1\}}\leq\frac{\mu_1}{\mu_1+\mu_2}\\
\mu_1/p_{\{1\}}, & p_{\{1\}}\geq\frac{\mu_1}{\mu_1+\mu_2},\\
\end{array}\right.$$
if  $(1-p_{\{2\}})/\mu_1>(1-p_{\{1\}})/\mu_2$. And, $$ \lambda^{R}=\mu_2/(1-p_{\{1\}})$$
if $(1-p_{\{2\}})/\mu_1=(1-p_{\{1\}})/\mu_2$. Similar to the $N$-model, the above can be obtained from Corollary~\ref{stab:cond}.
When $p_{\{1\}}= \mu_1/(\mu_1+\mu_2)$, maximum stability   $\lambda^{R}=\mu_1+\mu_2$ is obtained. 

\section{When does redundancy improve stability}
\label{sec:improve}

In this section, we compare the stability condition of the general redundancy system to that of the Bernoulli routing. 
Each job type has its own compatible servers, denoted by~$c$. Hence, given the compatible servers and the arrival rates of each type of jobs, we study whether redundancy can improve the stability condition. 

From Corollary~\ref{stab:cond}, it follows that  $\lambda^{R}=\min_{i=1,\ldots, i^*} CAR_i$. Together with~\eqref{stab:g-non}, we obtain the following   sufficient and necessary  conditions for redundancy to improve the stability condition.

\begin{corollary}\label{corol:imp-stab}
	The stability condition under redundancy is larger than under Bernoulli routing if and only if  $$
	\min_{i=1,\ldots,i^*, s\in\mathcal L_i}\{\frac{\mu_{s}}{\sum_{c\in\mathcal C_i(s)}p_c}\} \geq \min_{s\in S}\{\frac{\mu_{s}}{\sum_{c\in\mathcal C(s)} \frac{p_c}{\vert c \vert}}\}.$$
\end{corollary}


From inspecting the condition of Corollary~\ref{corol:imp-stab}, it is not clear upfront when redundancy would be better than Bernoulli. In the rest of the section, by applying Corollary~\ref{corol:imp-stab} to redundancy-$d$ and nested models, we will show that when the capacities of the servers are sufficiently heterogeneous, the stability of redundancy is larger than that of Bernoulli. In addition, numerical computations allow us to conclude that the degree of heterogeneity needed in the servers in order for  redundancy to be beneficial, decreases  in the number of servers, and increases in the number of redundant copies.


\subsection{Redundancy-$d$}
\label{sec:impr_d}

In this section, we compare the stability condition of the redundancy-$d$ model with homogeneous arrivals to that of Bernoulli routing. 
From~\eqref{stab:g-non}, we obtain that 
\begin{equation}\label{stab:non}\lambda^{B} = d\min_{i=1,\ldots,K} \left\{\frac{\mu_i}{\sum_{c\in\mathcal C(s)} p_c}\right\} =K\min_{i=1,\ldots,K} \mu_k.\end{equation}
From Corollary~\ref{corol:homo}
, we obtain that $\lambda^{R}=\min_{i=d,\ldots,K} \left\{ \frac{\binom{K}{d}}{\binom{i-1}{d-1}}\mu_i\right\}$. 
The following corollary is straightforward.

\begin{corollary}
	Let $\mu_1<\ldots<\mu_K$. The system under redundancy-$d$ and homogeneous arrivals has a strictly larger stability condition than the system under Bernoulli routing if and only if $$K\mu_1 <\min_{i=d,\ldots,K} \left\{ \frac{\binom{K}{d}}{\binom{i-1}{d-1}}\mu_i\right\}.$$
\end{corollary}

The following is straightforward, since   $\binom{i-1}{d-1}$ is increasing in $i$.

\begin{corollary}\label{corol:imp}
	Assume $\mu_1<\ldots<\mu_K$ and homogeneous arri-vals. The system under redundancy-$d$ has  a larger stability region than the Bernoulli routing if $\mu_1d<\mu_d$.   
\end{corollary}

Hence, if there exists a redundancy parameter $d$ such that $\mu_1d<\mu_d$, then adding $d$ redundant copies to the system improves its stability region. In that case, the stability condition of the system will improve by at least a factor~$\frac{\mu_d}{d\mu_1}$.  

%

In Table~\ref{table:stab_red_d}, we analyze how the heterogeneity of the server capacities impacts the stability of the system. 
We chose $\mu_k=\mu^{k-1}$, $k=1,\ldots,K$, so that the minimum capacity equals~1. Hence, for Bernoulli, $\lambda^{B}=K$. 
Under redundancy we have the following: 
For $\mu=1$ the system is a redundancy-$d$ system with homogeneous arrivals and server capacities, so that $\lambda^{R}= K/d$, \cite{Anton2019}. Thus, $\lambda^{R}<\lambda^{B}$ in that case. For $\mu>1$, that is, heterogeneous servers, we can apply Corollary~\ref{stab:cond} in order to find $\lambda^{R}$, that is, use Equation~\eqref{eq:lR}. More precisely, we create recursively the $i^*$ subsystems, calculate $CAR_i$ for each $i=1,\ldots, i^*$, so that  $\lambda^{R}=\min_{i=1\ldots, i^*}{CAR_i}$. We denote by $\mu^*$ the value of $\mu$ for which the stability region of the redundant system coincides with that of  Bernoulli routing, i.e., the value of $\mu$ such that $\lambda^{R}=\lambda^{B}$.
For $\mu<\mu^*$ (the area on the left-hand-side of the thick line in Table~\ref{table:stab_red_d}), Bernoulli has a larger stability region, while for $\mu>\mu^*$  (the area on the right-hand-side of the thick line in in Table~\ref{table:stab_red_d}), redundancy outperforms Bernoulli.

First, we observe that, for a fixed $d$, $\mu^*$ decreases as $K$ increases, and is always less than $\mu=2$. Therefore, as the number of servers increases, the level of heterogeneity that is needed in the servers in order to improve the stability under redundancy decreases.
Second, for fixed $K$, we also observe that $\mu^*$ increases as $d$ increases. This means that as the number of redundant copies $d$ increases, the server capacities need to be more heterogeneous in order to improve the stability region under redundancy. Finally,  focusing on the numbers in bold, we observe that when the number of servers $K$ is large enough   and the servers are heterogeneous enough (large   $\mu$), the stability region increases in the  number of redundant copies $d$. 


\begin{table}[h]
	\centering
	\caption{The maximum arrival rates $\lambda^{R}$ and $\lambda^{B}$ in a redundancy-$d$ system with homogeneous arrivals and capacities $\mu_k=\mu^{k-1}$.}
	\footnotesize
	\setlength{\tabcolsep}{2pt}
	\renewcommand{\arraystretch}{0.7}
	\begin{tabular}{|c|c|c|c|c|c|c|c|}
		\hline
		&	  & $\mu =1$ & $\mu=1.2$ & $\mu=1.4$ & $\mu=2$ & $\mu=3$ & $\mu^*$ \\
		\hline
		$K=3$ & Red-$2$   &\graya{1.5} &  \graya{2.16} &\Thickvrule{\graya{2.94}} &  6  & 9& 1.41\\
		& BR       & \graya{3} &  \graya{3} & \Thickvrule{\graya{3}}& 3 & 3 & \\ 
		\cline{1-4}\Cline{5-5}{1pt}\cline{6-8}	
		$K=4$ & Red-$2$   & \graya{2} & \Thickvrule{\graya{3.45}} & 5.48& 12 & 18&  1.26\\
		& BR       & \graya{4} &  \Thickvrule{\graya{4}} & 4& 4 & 4& \\ 
		\cline{1-3}\Cline{4-4}{1pt}\cline{5-8}	   
		$K=5$ & Red-$2$  &   \Thickvrule{\graya{2.5}}& 5.18 &   9.14 &  20&30&  1.19\\
		& BR       & \Thickvrule{\graya{5}} &5 & 5&5&  5& \\ 
		\hline
		$K=10$ & Red-$2$   & \Thickvrule{\graya{5}} &   22.39 &  41.16 &   90 & 135& 1.08\\
		& BR       & \Thickvrule{\graya{10}} & 10 & 10& 10 &10 & \\ 
		\hline
		\hline
		$K=4$ & Red-$3$   &  \graya{1.33} &  \graya{2.30} &  \Thickvrule{\graya{3.65}} &  10.66 &\textbf{36}&  1.44\\
		& BR       & \graya{4} & \graya{4} & \Thickvrule{\graya{4}}& 4 & 4& \\ 
		\cline{1-4}\Cline{5-5}{1pt}\cline{6-8}	   
		
		$K=5$ & Red-$3$   &  \graya{1.66} &  \Thickvrule{\graya{3.45}} & 6.40& \textbf{26.66}& \textbf{90}& 1.31\\
		& BR       & \graya{5}&  \Thickvrule{\graya{5}}& 5& 5 & 5 &\\ 
		\cline{1-3}\Cline{4-4}{1pt}\cline{5-8}	
		$K=10$ & Red-$3$   &	 \Thickvrule{\graya{3.33}}&  17.19 & \textbf{60.23}&  \textbf{320} &\textbf{1080} &1.13\\
		& BR       &  \Thickvrule{\graya{10}} & 10 & 10& 10 & 10 & \\ 
		\hline
	\end{tabular}
	\label{table:stab_red_d}
\end{table}

In Table~\ref{table:stab_red_d2}, we consider linearly increasing capacities on the interval $[1,M]$, that is $\mu_k= 1+\frac{M-1}{K-1}(k-1)$, for $k=1,\ldots,K$. In the area on the right-hand-side of the thick line, redundancy  outperforms Bernoulli. For this specific system, the following corollary is straightforward.

\begin{corollary}
	Under a redundancy-$d$ system with homogeneous arrivals and capacities $\mu_k=1+\frac{M-1}{K-1}(k-1)$, for $k=1,\ldots,K$, the redundancy system has stability condition: 
	$\lambda^{R} = \frac{M K }{d}$,
	for $d>1$, while $\lambda^{B}=K$.  Hence, the redundancy system outperforms the stability condition of the Bernoulli routing if and only if $M\geq d$.
\end{corollary}
Simple qualitative rules can be deduced. If $M\geq d$, redundancy is a factor $M/d$ better than Bernoulli.  Hence, increasing $M$, that is, the heterogeneity among the servers, is significantly beneficial for the redundancy system.   However,  the stability condition of the redundancy system degrades as the number of copies $d$ increases. 


\begin{table}[t]
	\centering
	\caption{The maximum arrival rates $\lambda^{R}$ and $\lambda^{B}$ in a redundancy-$d$ system with homogeneous arrivals and capacities $\mu_k=1+\frac{M-1}{K-1}(k-1)$.}
	\footnotesize
	\setlength{\tabcolsep}{2pt}
	\renewcommand{\arraystretch}{0.7}	
	\begin{tabular}{|c|c|c|c|c|c|c|}
		\hline
		&	  & $M=1$ & $M=2$ & $M=3$ & $M=4$ & $M=6$\\
		\hline
		$K=3$ & Red-$2$   &  \Thickvrule{\graya{1.5}} & 3 & 4.5&  6 & 9  \\
		& BR       & \Thickvrule{\graya{3}} & 3 & 3 & 3 & 3\\ 
		\hline
		$K=4$ & Red-$2$   &  \Thickvrule{\graya{2}}&  4& 6& 8& 12\\
		& BR       & \Thickvrule{\graya{4}}& 4 & 4& 4 & 4\\ 
		\hline	   
		
		$K=5$ & Red-$2$  & \Thickvrule{\graya{2.5}}&5&7.5& 10&15\\
		& BR       & \Thickvrule{\graya{5}} &5 & 5&5 &5\\ 
		\hline
		$K=10$ & Red-$2$   & \Thickvrule{\graya{5}}& 10& 15& 20&30\\
		& BR       & \Thickvrule{\graya{10}} & 10 & 10& 10 & 10\\ 
		\hline
		\hline
		$K=4$ & Red-$3$   & \graya{1.33}&\Thickvrule{\graya{2.66}}&  4& 5.33&  8\\
		& BR       & \graya{4} & \Thickvrule{\graya{4}} &4& 4 &4\\ 
		\hline	   
		
		$K=5$ & Red-$3$   &  \graya{1.66}& \Thickvrule{\graya{3.33}}& 5&6.66&  10\\
		& BR       & \graya{5}& \Thickvrule{\graya{5}}& 5& 5 &5 \\ 
		\hline
		$K=10$ & Red-$3$   &\graya{3.33}& \Thickvrule{\graya{6.66}}& 10&13.33&  20 \\
		& BR       & \graya{10} & \Thickvrule{\graya{10}} & 10& 10 &10 \\ 
		\hline
	\end{tabular}
	\label{table:stab_red_d2}
\end{table}

\subsection{Nested systems}
\subsubsection{$N$-model}
The stability condition of the $N$-model with Bernoulli routing is given by the following expression:
$$ \lambda^{B}= \left\{\begin{array}{ll}
2\min\{\mu_1,\mu_2\} ,& \mbox{ if } p=0\\
2\mu_1/(1-p), & \mbox{ if } 0 \leq p\leq \left(\frac{\mu_2-\mu_1}{\mu_1+\mu_2}\right)^+\\
2\mu_2/(1+p), &\mbox{ if } \left(\frac{\mu_2-\mu_1}{\mu_1+\mu_2}\right)^+<p\leq1.
\end{array}
\right.$$
The above set of conditions is obtained from the fact that under Bernoulli routing,  $\lambda^{B}=\min\{2\mu_1/(1-p) , \mu_2/(p+\frac{1}{2}(1-p))\}$. 
Note that $\lambda^{B}$ is a continuous function with a maximum $\mu_1+\mu_2$ at the point $p=\frac{\mu_2-\mu_1}{\mu_1+\mu_2}$.
Now, comparing $\lambda^{B}$ to $\lambda^{R}$ as  obtained in Section~\ref{sec:N}   leads to the following: 
\begin{corollary}
	Under an $N$-model, the stability condition under redundancy is larger than under Bernoulli routing under the following conditions: If $\mu_2\leq \mu_1$, then $p\in(\left(\frac{2\mu_2-\mu_1}{2\mu_2+\mu_1}\right)^+,1)$. If  $\mu_2>\mu_1$, then $p\in(0,(\frac{\mu_2-2\mu_1}{\mu_2})^+) \cup(
	\frac{2\mu_2-\mu_1}{2\mu_2+\mu_1},1)$.
\end{corollary}	

From the above we conclude that if $\mu_1$ is larger than $2\mu_2$, then redundancy is always better than Bernoulli, independent of the arrival rates of job types. For the case $\mu_2>\mu_1$, we observe that for $\mu_2$ large enough, redundancy will outperform Bernoulli.  
\subsubsection{$W$-based nested systems }
We consider the following structure of nested systems: $W$ (see Figure~\ref{fig:red_examples} c) ), $WW$ (Figure~\ref{fig:red_examples} d)) and $WWWW$. The latter is a $K=8$ server system that is composed of $2$ $WW$ models  and an additional job type~$c=\{1,\ldots,8\}$ for which all servers are compatible. For all  three models, we assume that a job is with probability $p_c=1/|\mathcal{C}|$ of type~$c$. 



In Table~\ref{table:stab_nested}, we analyze how  heterogeneity in the server capacities impacts the stability. First of all, note that $\lambda^{B}=K$. 
For redundancy, the value of $\lambda^{R}$ is given by~\eqref{eq:lR}, which  depends on the server capacities. In the table, we present these values for different values of the server capacities. In the upper part of the table, we let $\mu_k=\mu^{k-1}$ for $k=1,\ldots,K$.
We denote by $\mu^*$  the value of $\mu$ for which $\lambda^{R}=\lambda^{B}$. We observe that as the number of servers duplicate, the $\mu^*$ decreases, and is always smaller than 1.5. So that, as the number of servers increases, the level of heterogeneity that is needed  in order for redundancy to outperform Bernoulli decreases too. 

In the second part of the table we assume $\mu_k=1+\frac{M-1}{K-1}(k-1)$ for $k=1,\ldots,K$. 
We observe that when $M\geq K$ the stability condition under redundancy equals $\lambda^{R}=\vert\mathcal C\vert$, which is always larger than $\lambda^{B}=K$. However, as the number of servers increases, the maximum capacity of the servers, $M$,  needs to increase $M$ in order for redundancy to outperform Bernoulli.

\begin{table}[h]
	\centering
	\caption{The maximum arrival rates $\lambda^{R}$ and $\lambda^{B}$ in nested systems.}
	\footnotesize
	\setlength{\tabcolsep}{2pt}
	\renewcommand{\arraystretch}{0.65}
	\begin{tabular}{|c|c|c|c|c|c|c|}
		\hline
		$\mu_k=\mu^{k-1}$& & $\mu =1$ & $\mu=1.2$ & $\mu=1.4$ & $\mu=2$  &$\mu^*$ \\
		\hline
		$K=2$& $W$-model & \graya{1.5}&\Thickvrule{\graya{1.8}}&  2.10&  3& 1.33\\
		& BR       & \graya{2} & \Thickvrule{\graya{2}} &2& 2  & \\ 
		\cline{1-3}\Cline{4-4}{1pt}\cline{5-7}	   
		$K=4$&$WW$-model & \Thickvrule{\graya{2.33}} &   4.03&  4.90&  7 & 1.19 \\
		& BR       & \Thickvrule{\graya{4}} & 4 & 4& 4& \\ 
		\hline
		$K=8$& $WWWW$-model  &\Thickvrule{\graya{3.75}}&8.64 &10.5& 15& 1.17\\
		& BR   &\Thickvrule{\graya{8}}  & 8& 8 &8& \\ 
		\hline
		\hline
		$\mu_k=1+\frac{M-1}{K-1}(k-1)$& & $M=1$ & $M=2$ & $M=4$ & $M=6$ & $M=8$  \\
		\hline
		$K=2$& $W$-model  &\Thickvrule{\graya{1.5}} & 3 & 3 & 3 & 3\\
		&BR 		& \Thickvrule{\graya{2}} & 2 & 2 & 2 & 2 \\ 
		\hline
		$K=4$&$WW$-model  &\Thickvrule{\graya{2.33}} & 4.66 &   7 &  7 & 7 \\
		& BR   & \Thickvrule{\graya{4}}   & 4 & 4 & 4 & 4  \\
		\cline{1-3}\Cline{4-4}{1pt}\cline{5-7}
		$K=8$&$WWWW$-model  &\graya{3.75}&\Thickvrule{\graya{7.14}} &10.71& 12.85& 15 \\
		&BR   &\graya{8}  & \Thickvrule{\graya{8}}& 8 &8&8  \\ 
		\hline
	\end{tabular}
	
	\label{table:stab_nested}
\end{table}

\section{Proof of Proposition~\ref{rem:stab}}
\label{sec:stab_cond_proof}
In this section, we prove that the condition in Proposition~\ref{rem:stab} is sufficient and necessary for the respective subsystem to be stable. As we observe in Section~\ref{sec:example}, there are two main issues concerning the evolution of redundancy systems with heterogeneous capacities. First of all, the number of copies in a particular server decreases, only if a certain subset of servers is already in steady state. Secondly, for a particular server $s\in S$, the instantaneous departure of that server might be larger than $\mu_s$ due to copies leaving in servers other than $s$. This makes the dynamics of the system complex. In order to prove Proposition~\ref{rem:stab},  we therefore construct  upper  and lower bounds of our system for which  the dynamics   are  easier to characterize. Proving that the upper bound (lower bound) is stable (unstable) directly implies that the original system is also stable (unstable).  This will be done in Proposition~\ref{prop:dir1} and Proposition~\ref{prop:dir2}. All proofs of this section can be found in Appendix B.

\subsection*{Sufficient stability condition}
\label{sec:suf}

We define the Upper Bound ($UB$) system as follows.
Upon arrival, each job is with probability $p_c$ of type $c$ and sends identical copies to all servers $s\in c$. In the UB system, a type-$c$ job departs the system  {\bf only when all copies in the set of servers $\mathcal R(c)$ are fully served.} We recall that the set $\mathcal{R}(c)$  denotes the set of servers where a type-$c$ job achieves maximum capacity-to-fraction-of-arrivals ratio. When this happens, the remaining copies that are still in service (necessarily not in a server in  $\mathcal R(c)$) are immediately removed from the system. We denote by $N_c^{UB}(t)$ the number of type-$c$ jobs present in the UB system at time~$t$. 

We note that the UB system is closely related to the one in which  copies of type-$c$ jobs are only  sent to servers in $\mathcal R(c)$. However, the latter system is of no use for our purposes  as it is  neither an upper bound nor a lower bound of the original system.

We can now  show the first implication of Proposition~\ref{rem:stab}, that is, we prove that $\lambda<CAR_l$, for all $l=1,\ldots, i$, implies stability  of the servers in the set $\mathcal{L}_i$. We do this by analyzing the   UB system for which stability of the servers $\mathcal{L}_i$ follows intuitively as follows:  Given a server  $s\in \mathcal{L}_1$ and any  type $c\in C(s)$, it holds that  $\mathcal R(c)\subseteq \mathcal{L}_1(c)$. Hence, a  server in $\mathcal{L}_1$ will need to fully serve all arriving copies. Therefore each server $s$, with $s\in \mathcal{L}_1$, behaves as an {M/G/1 PS} queue, which is stable if and only if its arrival rate of copies, $\lambda \sum_{c\in C_1(s)}p_c$, is strictly smaller than its departure rate, $\mu_s$.  Assume now that for all  $l=1,\ldots,i-1$   the subsystems $S_l$ are stable and we want to show that servers in $\mathcal{L}_i$ are stable as well. First of all, note that in the fluid limit, all types $c$ that do not exist in the $S_i$-subsystem, i.e., $c\notin C_i(s)$, will after a finite amount of time  equal (and remain) zero, since they are stable. For the remaining types $c$ that have copies in server $s\in \mathcal{L}_i$, i.e., $s\in c$ with $s\in\mathcal L_i$, it will hold that their servers with maximum capacity-to-fraction-of-arrivals ratio are  $\mathcal{R}(c)\subseteq \mathcal{L}_i$.  Due to the characteristics of the upper-bound system, all copies sent to these servers will need to be served. Hence, a server~$s\in\mathcal{L}_i$ behaves in the fluid limit as an  M/G/1 PS queue with arrival rate $\lambda \sum_{c\in C_i(s)} p_c$ and departure rate $\mu_s$. In particular, such a queue is stable if and only if $\lambda \sum_{c\in C_i(s)} p_c < \mu_s$.

\begin{proposition}
	\label{prop:suff}
	For $i\leq i^*$, the set of servers $s\in\mathcal L_i$ in the $UB$ system is stable if $\lambda < CAR _l, \textrm{  for all  } l=1,\ldots,i.$
\end{proposition}

In the following, we prove that  $UB$  provides an upper bound on the original system. To do so, we show that every job departs earlier in the original system than in the $UB$ system. In the statement, we assume that in case a job has already departed in the original system, but not in the UB system, then its attained service in all its servers in the original system is set equal to its service requirement~$b_{cj}$. 

\begin{proposition}\label{prop:ub}
	Assume  $N_c(0)=N^{UB}_c(0)$  and $a_{cjs}(0)=a^{UB}_{cjs}(0)$, for all $c,j,s$. Then, $N_c(t)\leq N^{UB}_c(t)$ and $a_{cjs}(t)\geq a^{UB}_{cjs}(t)$, for all $c,j,s$ and $t\geq 0$. 	
\end{proposition}

Together with Proposition~\ref{prop:suff}, we obtain the following result for the original system.  
\begin{proposition}
	\label{prop:dir1}
	For a given $i\leq i^*$, servers $s\in \mathcal{L}_i$ are stable if $\lambda<CAR_l$, for all $l=1,\ldots, i$. 
\end{proposition}

\begin{remark}
	In \cite{Anton2019}, the authors show that for the redundancy-$d$ system with homogeneous arrivals and server capacities, the system where all the copies need to be served is an upper bound.  We note that this upper bound coincides with our upper bound (in that case  $\mathcal L_1=S$).
	Nevertheless, the proof approach is different. In \cite{Anton2019}, see also \cite{Raaijmakers2019}, the proof followed directly, as each server in the upper bound system behaved as an M/G/1 PS queue. 
	In the heterogeneous server setting studied here, the latter is no longer true. Instead, it does apply recursively when considering the fluid regime: In order to see a server as a PS queue in the fluid regime, one first needs to argue that the types that have copies in higher capacity-to-fraction-of-arrivals servers are  $0$ at a fluid scale.
\end{remark}

\begin{remark}
	We note that the light-tail assumption on the service time distribution, see Section~\ref{sec:model},  is an assumption needed in order to prove Lemma \ref{lemma:Harris} (see Appendix~B for more details).
	
\end{remark}

\subsection*{Necessary stability condition}
\label{sec:nec}
In this section we prove the necessary stability condition of Proposition~\ref{rem:stab}. 
Let us first define
$$\iota:=\min\left\{l=1,\ldots,i^* : \lambda>CAR_l \right\}.$$ 
We note that for any $i<\iota$, $\lambda<CAR_i$. Hence, the servers in $\mathcal L_i$, with $i<\iota$ are stable, see Proposition~\ref{prop:suff}. We are left to prove that 
the servers in $S_\iota$ cannot be stable.
In order to do so, we construct a lower-bound system. 

In the $S_\iota$ subsystem, the capacity-to-fraction-of-arrivals ratios are such that for all $s\in S_\iota$,
$\mu_s/(\sum_{c\in\mathcal C_\iota(s)}p_c) \leq CAR_\iota$. We will construct a lower bound (LB) system in which the resulting capacity-to-fraction-of-arrivals ratio is $CAR_\iota$ for all servers $s\in S_\iota$.
We use the superscript LB in the notation to refer to this system, which is defined as follows. First of all, we only want to focus on the $\mathcal{S}_\iota$ system, hence, we set the arrival rate $p^{LB}_c=0$ for types $c\in\mathcal C\backslash\mathcal C_\iota$, whereas the arrival rate for types $c\in\mathcal C_\iota$ remain unchanged, i.e.,  $p_c^{LB}=p_c$. 
The capacity of servers 
$s\in S_\iota$ in the LB-system is set to 
$$\mu_s^{LB}:=\mu_{\tilde s}\frac{\sum_{c\in\mathcal C_\iota(s)} p_c}{\sum_{c\in\mathcal C_\iota(\tilde s)} p_c}
=
CAR_\iota\cdot(\sum_{c\in\mathcal C_\iota(s)} p_c),$$ 
where  $\tilde s\in \mathcal{L}_\iota$. 
Additionally, in the LB-system, we assume that each copy of a type-$c$ job receives the same amount of capacity, which is equal to the highest value of $\mu_s^{LB}/M_s^{LB}(t)$, $s\in c$.
We therefore  define the service rate for a job of type $c$ by 
\begin{equation}
\label{eq:phi}
\phi_c^{LB}(\vec N^{LB}(t)):=
\max_{s\in c}\left\{\frac{\mu_s^{LB}}{M_s^{LB}(t)}\right\},
\end{equation}
where $c\in\mathcal C_\iota$ (instead of $\phi_s(\cdot)$ for a copy in server $s$ in the original system).    
The cumulative amount of capacity that a type-$c$ job receives is 
$$\eta^{LB}_c(v,t):=\int_{x=v}^{t} \phi^{LB}_c(\vec N^{LB}(x)) dx,  \mbox{ for  }c\in\mathcal C_\iota. $$

\begin{proposition}
	\label{corol:LB}
	In the LB-system, the set of servers $s\in  S_\iota$   is unstable if  $\lambda > CAR_\iota$.
\end{proposition}

We now prove  that LB is a lower bound for the original system. 

\begin{proposition}\label{prop1} 
	Assume $N_c(0)=N^{LB}_c(0)$, for all $c$. Then, 
	$N_c(t)\geq_{st} N^{LB}_c(t)$, for all $c\in \mathcal C$ and $t\geq0$.
\end{proposition}

Combining Proposition~\ref{corol:LB} with Proposition~\ref{prop1}, we obtain the following result for the original system.
\begin{proposition}
	\label{prop:dir2}
	Servers $s\in S_i$  are unstable if  there is an $l=1,\ldots, i$ such that  $\lambda > CAR_l$.
\end{proposition}

\begin{remark}
	In the special case of redundancy-$d$ with homogeneous arrivals and server capacities, \cite{Anton2019}  used a lower bound that consisted in modifying the service rate obtained per job type, as in~\eqref{eq:phi}.  This lower bound coincides with our lower bound, since  with homogeneous arrivals and servers  it holds that $\mu_s^{LB}=\mu_s=\mu$.
	The difficulty when studying heterogeneous servers in a general redundancy structure, as we do in this paper, lies in the fact that the load received in each server is different. In order to show that the fluid limit of the server with the minimum number of copies is increasing (in the lower bound), we need to adequately modify the server capacities in order to make sure that the capacity-to-fraction-of-arrival rates in each of the servers is equal. 

\end{remark}
\section{Numerical analysis}
\label{sec:numerics}
We have implemented a simulator in order to assess the  impact of redundancy. In particular, we evaluate the following: 
\begin{itemize}
	\item For PS servers, we numerically compare the performance of redundancy with   Bernoulli routing (in Section~6  this was done analytically for the stability conditions). 
	\item We compare redundancy to  the Join the Shortest Queue (JSQ) policy according to which each job is dispatched to the compatible server that has the least number of jobs (ties are broken at random).   In a recent paper,~\cite{cruise2020stability}, it was shown  that JSQ -- with exponential service time distributions -- combined with size-unaware scheduling disciplines such as FCFS, ROS or PS, is maximum stable, i.e., if there exists a static dispatching policy that achieves stability, so will JSQ. 
	\item We compare the performance between PS,  FCFS and  Random Order of Service (ROS), when the service time distribution is exponential and bounded Pareto. 
\end{itemize}
Our  simulations consider a  large number of busy periods $(10^6)$, so that the variance and confidence intervals of the mean number of jobs in the system are sufficiently small. 

\begin{figure}[b]
	\begin{subfigure}{.32\textwidth}
		\centering
		\includegraphics[width=1.1\textwidth]{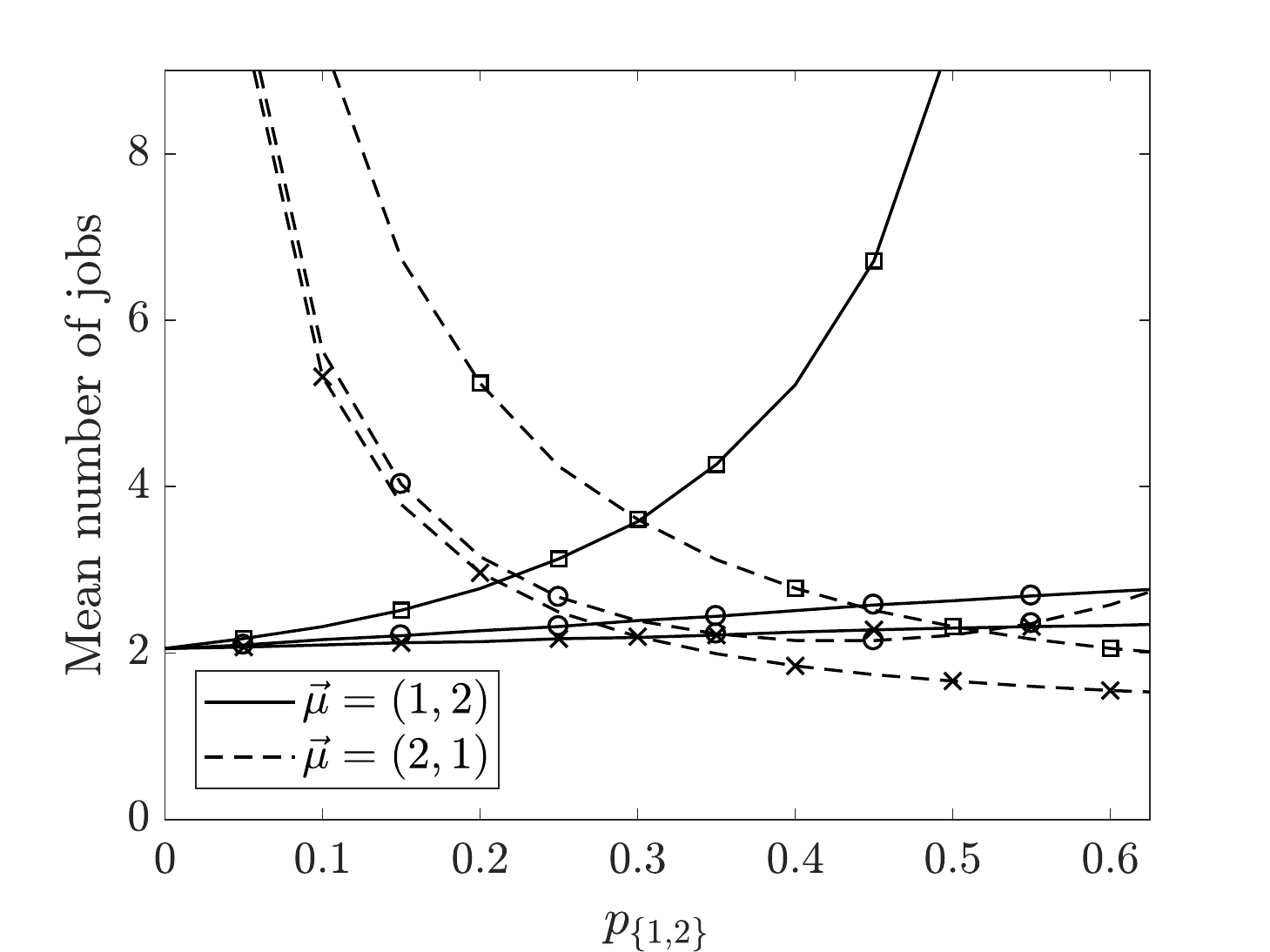}\\
		a) $\lambda=1.5$
	\end{subfigure}
	\begin{subfigure}{.32\textwidth}
		\centering
		\includegraphics[width=1.1\textwidth]{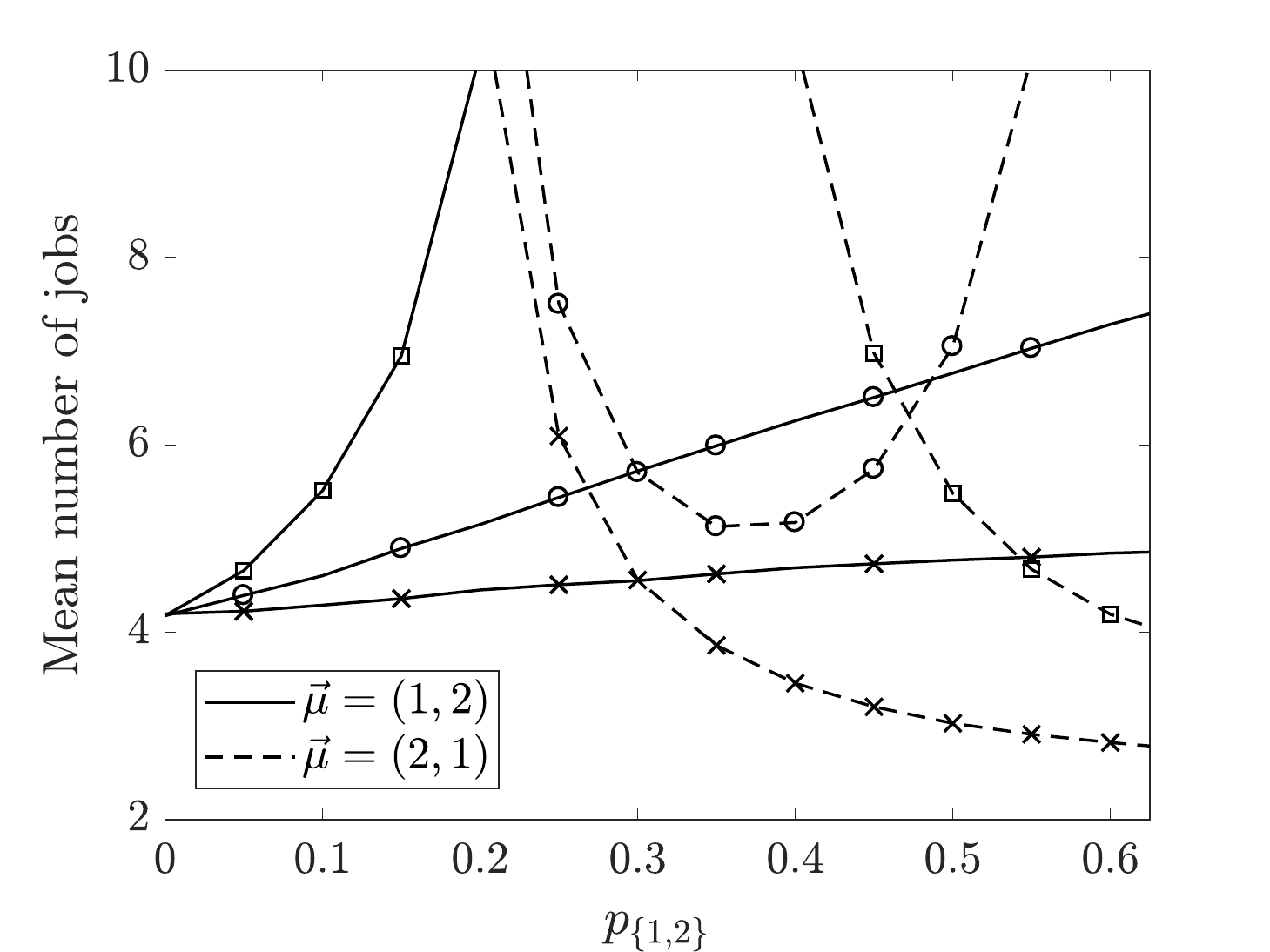}\\
		b) $\lambda=2$
	\end{subfigure}
	\begin{subfigure}{.32\textwidth}
		\centering
		\includegraphics[width=1.1\textwidth]{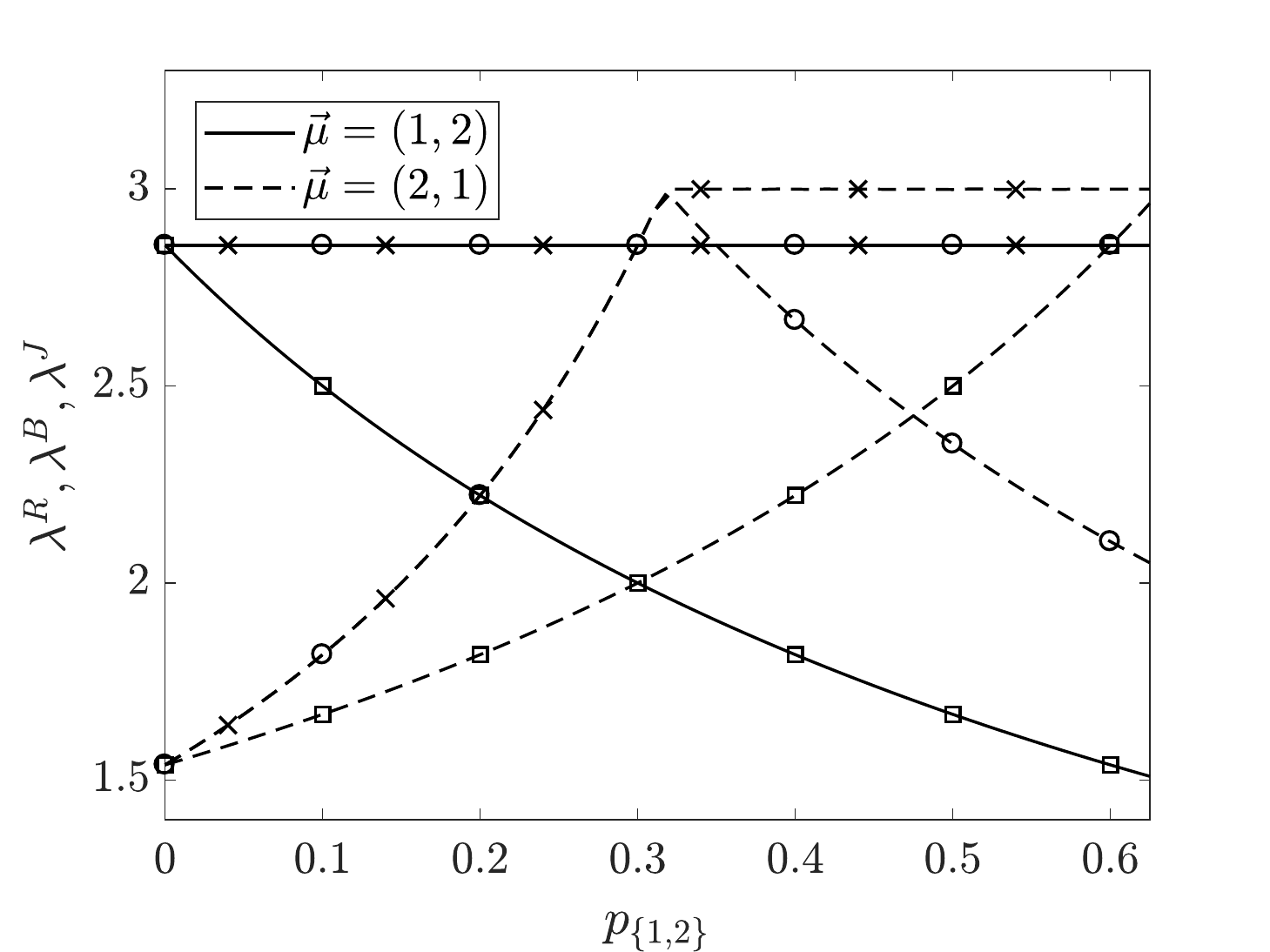}\\
		c)
	\end{subfigure}
	\caption{$W$-model with $p_{\{1\}}=0.35$, $p_{\{2\}} = 1- p_{\{1\}}-p_{\{1,2\}}$. 
		{a) and b) depict the mean number of jobs under redundancy with PS ($\circ$), Bernoulli routing ($\square$) and JSQ ($\times$) for $\lambda=1.5$ and $\lambda=2$. c) depicts the stability regions $\lambda^{R}$, $\lambda^{B}$ and $\lambda^J$.}}
	\label{Fig:W_exp1}
\end{figure}

\ {\bf Exponential service time distributions:}
In Figure~\ref{Fig:W_exp1} we consider the $W$-model with exponential service time distributions. We set  $p_{\{1\}}=0.35$ and $p_{\{2\}} + p_{\{1,2\}} = 0.65$, and vary the value of  $p_{\{1,2\}}$. We consider either $\vec\mu=(1,2)$ or $\vec\mu=(2,1)$,
The only redundant job type is $\{1,2\}$, thus as $p_{\{1,2\}}$ increases, we can observe  how increasing the fraction of redundant jobs affects the performance. We also note that when $p_{\{1,2\}}$ increases, the load in server 1 increases as well,  whereas the load in server 2 stays constant. In Figure~\ref{Fig:W_exp1} a) and b) we depict the mean number of jobs under  redundancy, Bernoulli routing and JSQ when the server policy is PS. In Figure~\ref{Fig:W_exp1} c) we plot $\lambda^{R}$, $\lambda^{B}$ and $\lambda^J$ using the analysis of Section 5.2.2. and \cite{cruise2020stability}, respectively.

We observe from Figure~\ref{Fig:W_exp1}~a)~and~b) that when $\vec \mu=(1,2)$, redundancy performs better than Bernoulli routing. This difference becomes larger as  $p_{\{1,2\}}$ increases. This is due to the fact that 
the redundancy policy does better in exploiting the larger capacity of server 2 than Bernoulli, which becomes more important as $p_{\{1,2\}}$ increases.
In addition, we note that for redundancy, Bernoulli and JSQ, the mean number of jobs increases as  $p_{\{1,2\}}$ increases. The reason for this is that as $p_{\{1,2\}}$ increases, the load on server~1 increases. Since server~1 is the slow server, this increases the mean number of jobs.

In the opposite case, i.e., $\vec \mu=(2,1)$, the mean number of jobs is non-increasing in $p_{\{1,2\}}$. This is because  as $p_{\{1,2\}}$ increases, the load on server~1 increases. Since server~1 is now the  fast server, this has a positive effect on the performance (decreasing mean number of jobs). However, as $p_{\{1,2\}}$ gets larger, the additional load (created by the copies) makes that the performance  can be  negatively impacted. This happens for $\lambda=2$, where the mean number of jobs under redundancy is a U-shape function. 
We furthermore observe that in the $\vec \mu=(2,1)$ case, redundancy outperforms   Bernoulli for any value of $p_{\{1,2\}}$ when $\lambda=1.5$. However, when $\lambda=2$, Bernoulli outperforms redundancy when $p_{\{1,2\}}>0.49$. This is due to the  additional load, generated under redundancy, that becomes more pronounced as $p_{\{1,2\}}$ becomes larger.

We also observe in Figure~\ref{Fig:W_exp1}, that under both $\vec \mu=(1,2)$ and $\vec \mu=(2,1)$,  JSQ outperforms redundancy. For small values of $p_{\{1,2\}}$ the difference is rather small, however it becomes larger as $p_{\{1,2\}}$  increases due to the additional load that redundancy creates.   
However that this improvement does not come for free, as JSQ requires precise information of the queue lengths at all times.

In  Figure~\ref{Fig:W_exp1} c), we observe that redundancy consistently has a larger stability region than Bernoulli  in the $\vec\mu=(1,2)$ case and  for $p_{\{1,2\}}\in[0,0.5)$ in the $\vec\mu=(2,1)$ case. 
We let  $\lambda^J$ be the value of $\lambda$ such that JSQ is stable if $\lambda<\lambda^J$ and unstable if $\lambda>\lambda^J$. Using \cite{cruise2020stability}, $$\lambda^J = \max_{p_{c,s} \ge 0, \sum_{s} p_{c,s}=p_c} \min_{s} { \mu_s \over \sum_{c} p_{c,s} }.$$  
We observe that  the stability condition under redundancy coincides on a large region with that of JSQ, which, in  view of the results of~\cite{cruise2020stability},   implies that redundancy is in that region maximum stable. 

\begin{figure}[b]
	\centering
	\begin{subfigure}{.32\textwidth}
		\centering
		\includegraphics[width=1\textwidth]{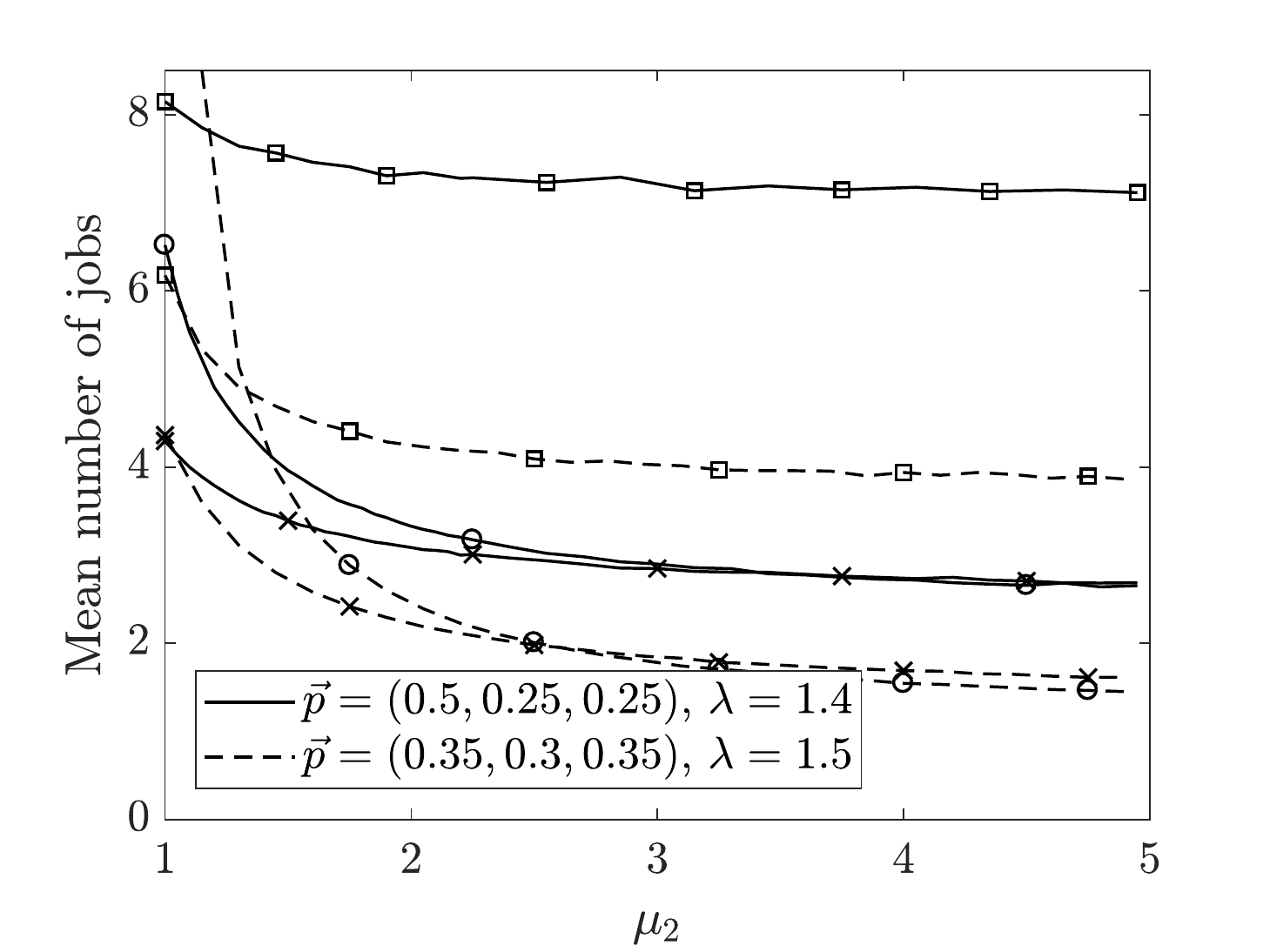}\\
		a) 
	\end{subfigure}
	\begin{subfigure}{.32\textwidth}
		\centering
		\includegraphics[width=1\textwidth]{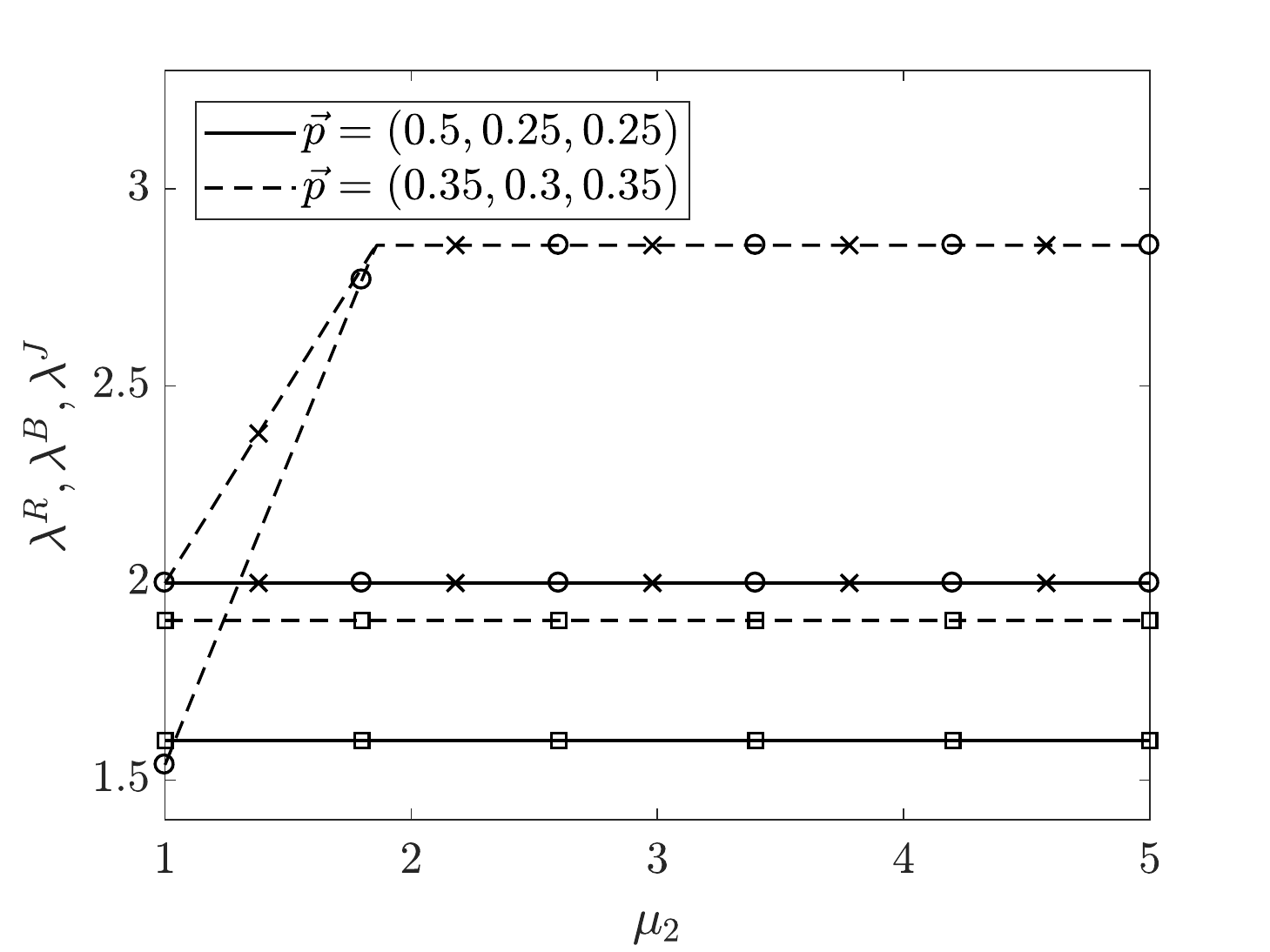}\\
		b) 
	\end{subfigure}
	\caption{$W$-model with fixed parameters $\vec p$ and $\mu_1=1$: a) depicts  the mean number of jobs under redundancy ($\circ$), Bernoulli routing ($\square$) and JSQ ($\times$), and  b) depicts the stability regions $\lambda^{R}$, $\lambda^{B}$ and $\lambda^J$.}
	\label{Fig:W_exp3}
\end{figure}

\begin{figure}[t]
	\centering
	\begin{subfigure}{.32\textwidth}
		\centering
		\includegraphics[width=0.9\textwidth]{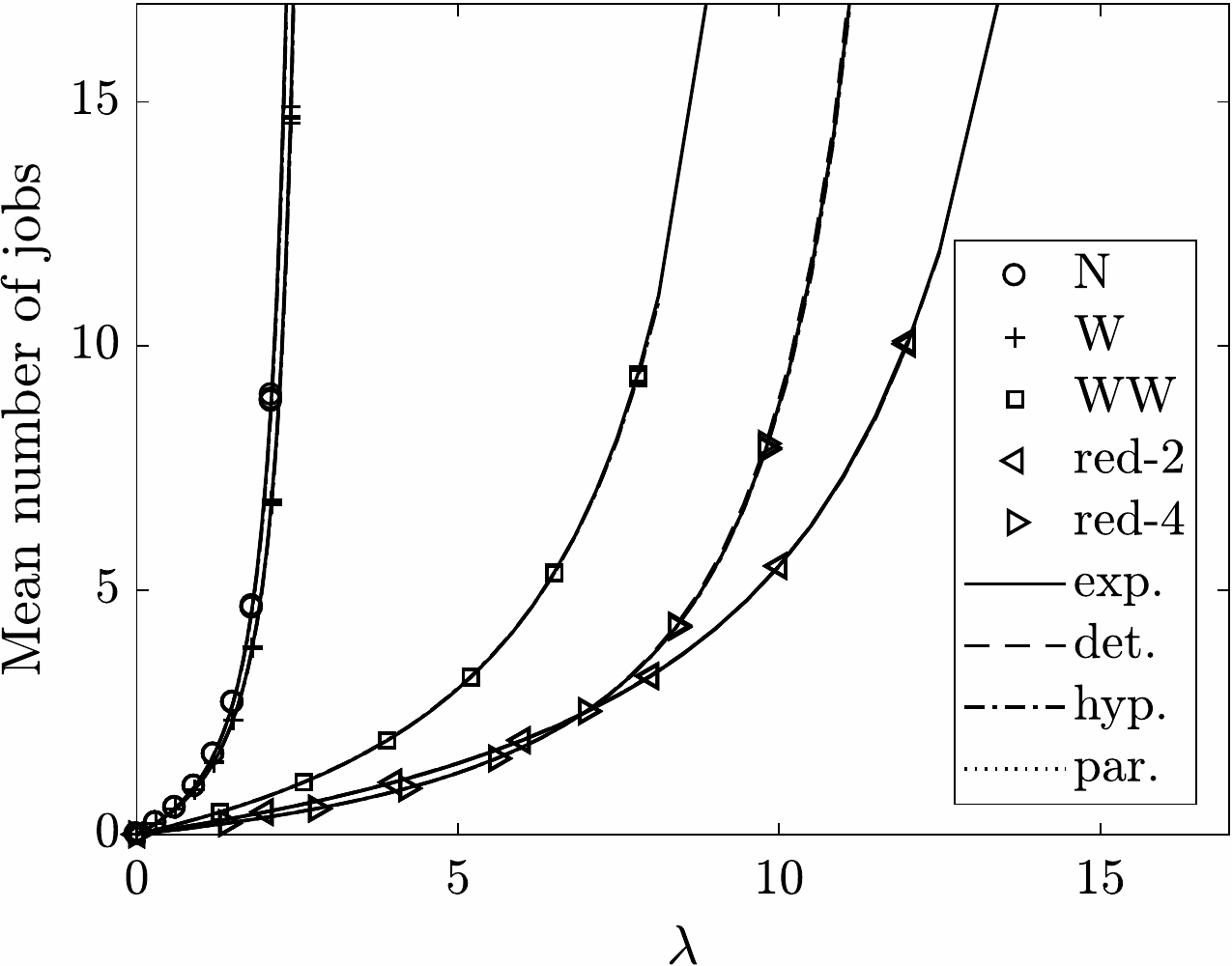}\\
		a)
	\end{subfigure}
	\begin{subfigure}{.32\textwidth}
		\centering
		\includegraphics[width=1.05\textwidth]{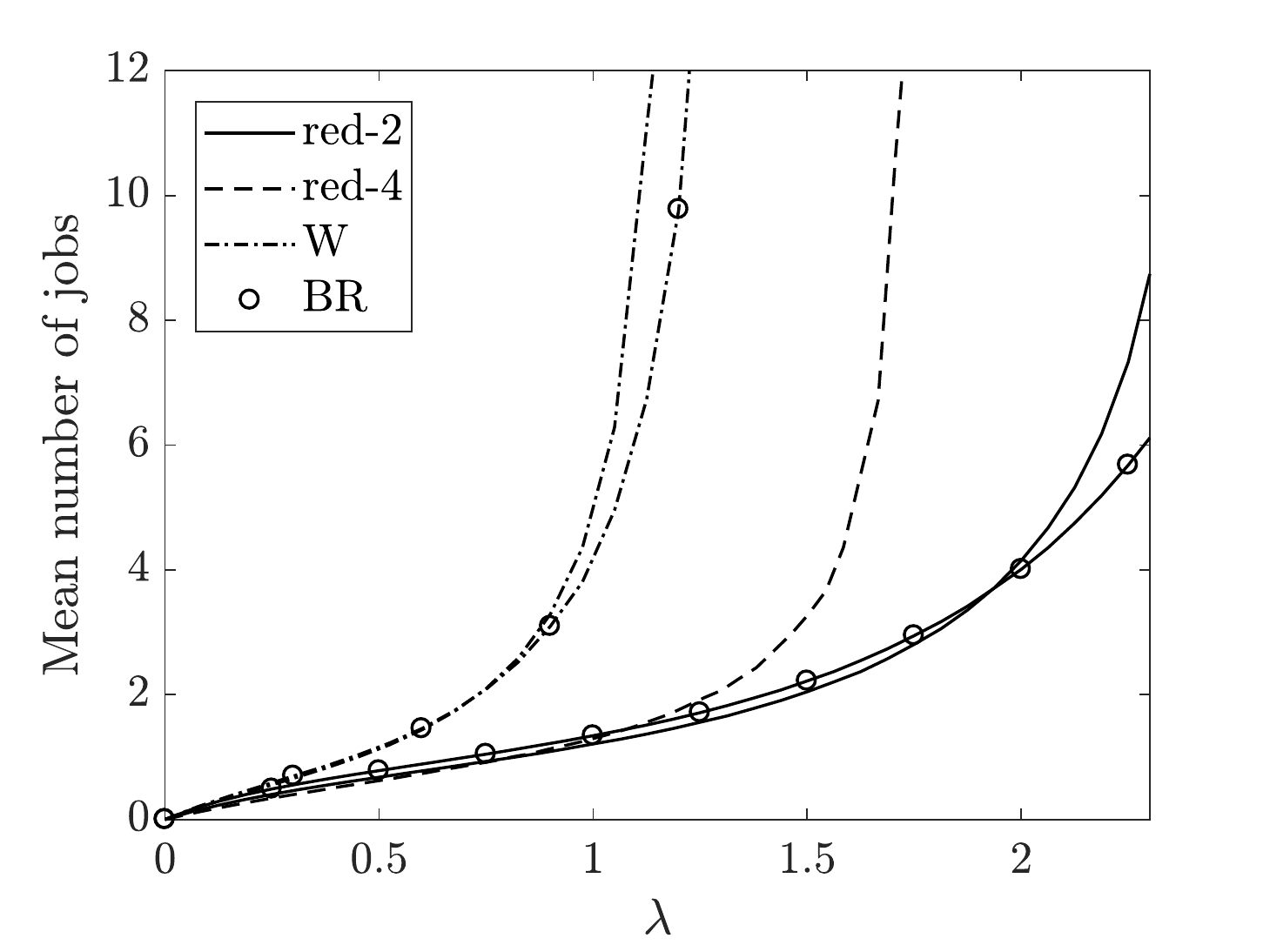}\\
		b) 
	\end{subfigure}
	\caption{Mean number of jobs in the system with respect to $\lambda$: a) Non-exponential service times and models $N$, $W$, $WW$, and  redundancy-$2$ ($K=5$), and redundancy-$4$ ($K=5$) models. We chose  $\vec\mu=(1, 2)$ for the  $N$ and $W$ model,    $\vec\mu=(1, 2, 4, 6)$ for the $WW$ model, and $\vec\mu=(1, 2, 4, 6, 8)$ for redundancy-$d$. b) Markov modulated server capacities in the $W$, and  redundancy-$2$ ($K=5$), and redundancy-$4$ ($K=5$) models.}
	\label{Fig2}
\end{figure}

In Figure~\ref{Fig:W_exp3} we simulate the performance of the $W$ model for different values of $\mu_2$, while keeping  fixed $\vec p=(p_{\{1\}},p_{\{2\}},p_{\{1,2\}})$ and $\mu_1 =1 $.  In Figure~\ref{Fig:W_exp3}~a) we plot the mean number of jobs and we see that for both configurations of $\vec p$, the performance of the redundancy with PS, Bernoulli and JSQ improve as $\mu_2$ increases. The gap between redundancy and Bernoulli is significant in both cases. The reason can be deduced from Figure~\ref{Fig:W_exp3}~b), where we plot $\lambda^{R}$, $\lambda^{B}$, and $\lambda^J$,  with respect to $\mu_2$. We observe in Figure~\ref{Fig:W_exp3}~a) that redundancy and JSQ converge to the same performance as $\mu_2$ grows large. Intuitively, we can explain this by observing that for very large values of $\mu_2$, with both redundancy and JSQ,  all jobs of type $p_{\{1,2\}}$ get served in server 2. 
We observe in Figure~\ref{Fig:W_exp3}~b) that the stability conditions with redundancy and JSQ are very similar.

\  {\bf General service time distributions:}
In Figure~\ref{Fig2} a) we investigate the performance of redundancy with PS for several non-exponential distributions. In particular, we consider the following distributions for the service times: deterministic, hyperexponential, and bounded Pareto. With the hyperexponential distribution,  job sizes are exponentially distributed with parameter $\mu_1$ ($\mu_2$) with probability $q$ ($1-q$). For Pareto the density function is $\frac{1- (k/x)^\alpha }{ (1- (k/\tilde q)^\alpha)}$, for $k \leq x \leq \tilde q$. We choose the parameters so that the mean service time equals~1. Namely for the hyperexponential distribution parameters are $q=0.2$, $\mu_1=0.4$ and $\mu_2=1.6$, and for the bounded Pareto distribution are $\alpha=0.5$, $\tilde q=6$ and $k=1/\tilde q$. In Figure~\ref{Fig2} a), we plot the mean number of jobs as a function of $\lambda$ for the $N$, $W$, $WW$, and redundancy-$2$ ($K=5$), and redundancy-$4$  ($K=5$) models. The respective parameters $\vec p$ are chosen such that the system is stable for the simulated arrival rates. We observe that for the five systems, performance seems to be nearly insensitive to the service time distribution, beyond its mean value.




\  {\bf Markov-modulated capacities:}
In Figure~\ref{Fig2} b) we consider a variation of our model where servers' capacities fluctuate over time. More precisely, we assume that each server has an exponential clock, with mean $\epsilon$. Every time the clock rings, the server samples a new value for $S$ from Dolly(1,12), see Table~\ref{tabdolly} and sets its capacity equal to $1/S$. The Dolly(1,12) distribution is a 12-valued discrete distribution  that was empirically obtained by analyzing traces in Facebook and Microsoft clusters, see  \cite{Ananthanarayanan13,Gardner17b}.

In Figure~\ref{Fig2} b) we plot the mean number of jobs for a $K=5$ server system with  redundancy-2 and redundancy-4, and for the  $W$-model under redundancy, and we compare it with   Bernoulli routing. Arrival rates are equal for all classes. It can be seen that with Bernoulli routing, both redundancy-2 and redundancy-4 become equivalent systems, and hence their respective curves overlap. { The general observation is that in this setting with identical servers, Bernoulli routing performs better than redundancy.} 
Further research is needed to understand whether with heterogeneous Markov-modulated servers, redundancy can be beneficial.
\begin{table}[ht]
	\footnotesize
	\centering
	\setlength{\tabcolsep}{3pt}
	\renewcommand{\arraystretch}{0.85}
	\caption{The Dolly(1,12) empirical distribution for the slowdown \cite{Ananthanarayanan13}. The capacity is set to $1/S$.}
	\begin{tabular}{|c|c|c|c|c|c|c|c|c|c|c|c|c| }	
		\hline
		$S$ & 1 & 2 & 3 & 4 & 5 & 6 & 7 & 8 & 9 & 10 & 11 & 12 \\
		\hline 
		Prob & 0.23 & 0.14 & 0.09 & 0.03 & 0.08 & 0.10 & 0.04 & 0.14 & 0.12  &   0.021 &  0.007 &  0.002 \\
		\hline
	\end{tabular}
	
	\label{tabdolly}
\end{table}



\textbf{FCFS and ROS scheduling discipline:}
The stability condition under  FCFS or ROS   and identical copies is not known. An exception is the redundancy-$d$ model with  homogeneous arrivals and server capacities for which \cite{Anton2019} characterizes the stability condition under ROS, FCFS and PS. There it was shown that   ROS is maximum stable, i.e., the stability condition is $\lambda <\mu K$, and that under FCFS the stability condition is $\lambda<\bar{\ell} \mu$, where $\bar\ell$ is the mean number of jobs in service in a so-called associated  saturated system. In addition, it was shown that for this specific setting,   the stability region under PS is smaller than under FCFS and ROS.

\begin{figure}[t]
	\begin{subfigure}{.32\textwidth}
		\centering
		\includegraphics[width=1.1\textwidth]{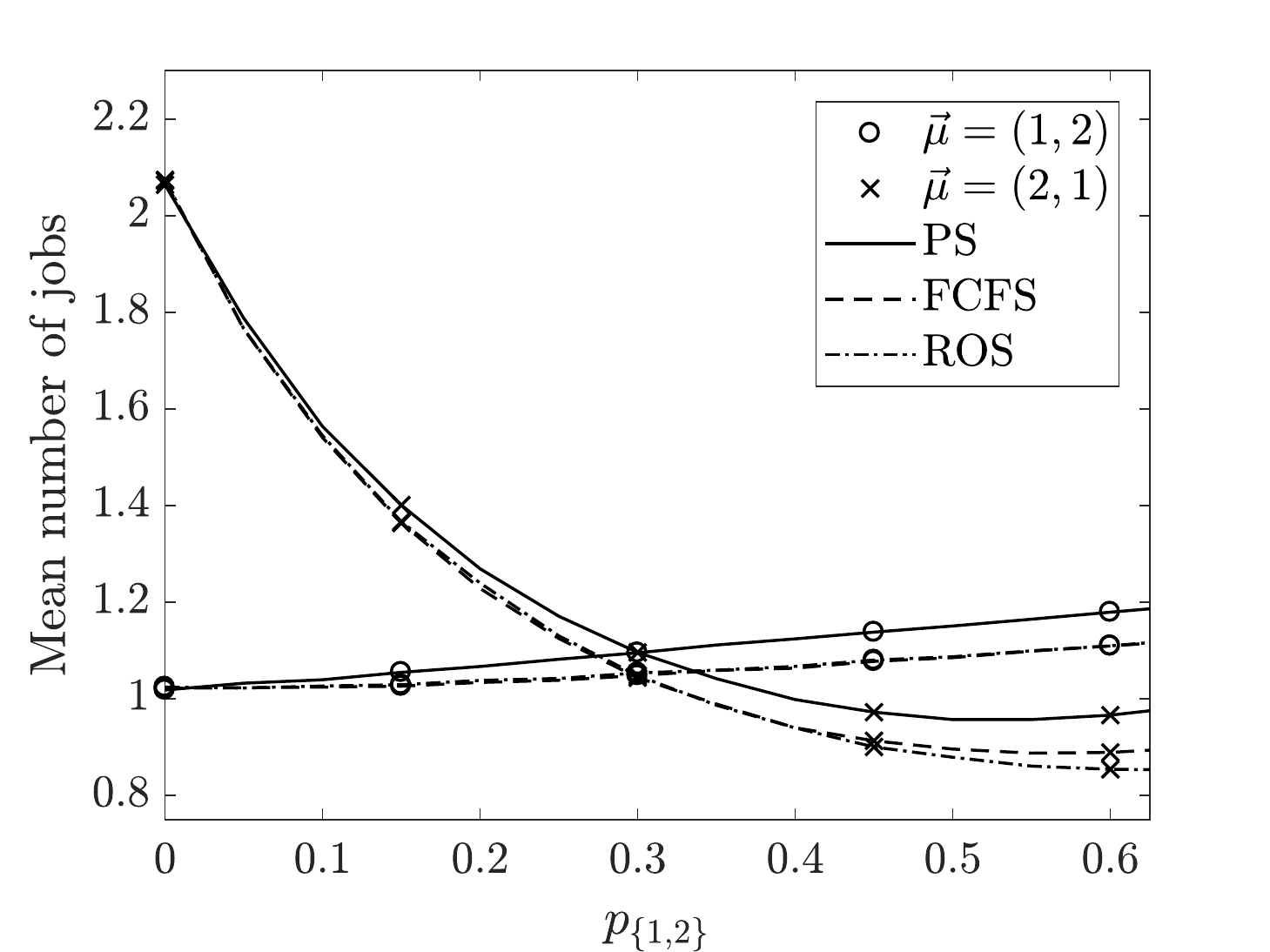}\\
		a) $\lambda=1$
	\end{subfigure}
	\begin{subfigure}{.32\textwidth}
		\centering
		\includegraphics[width=1.1\textwidth]{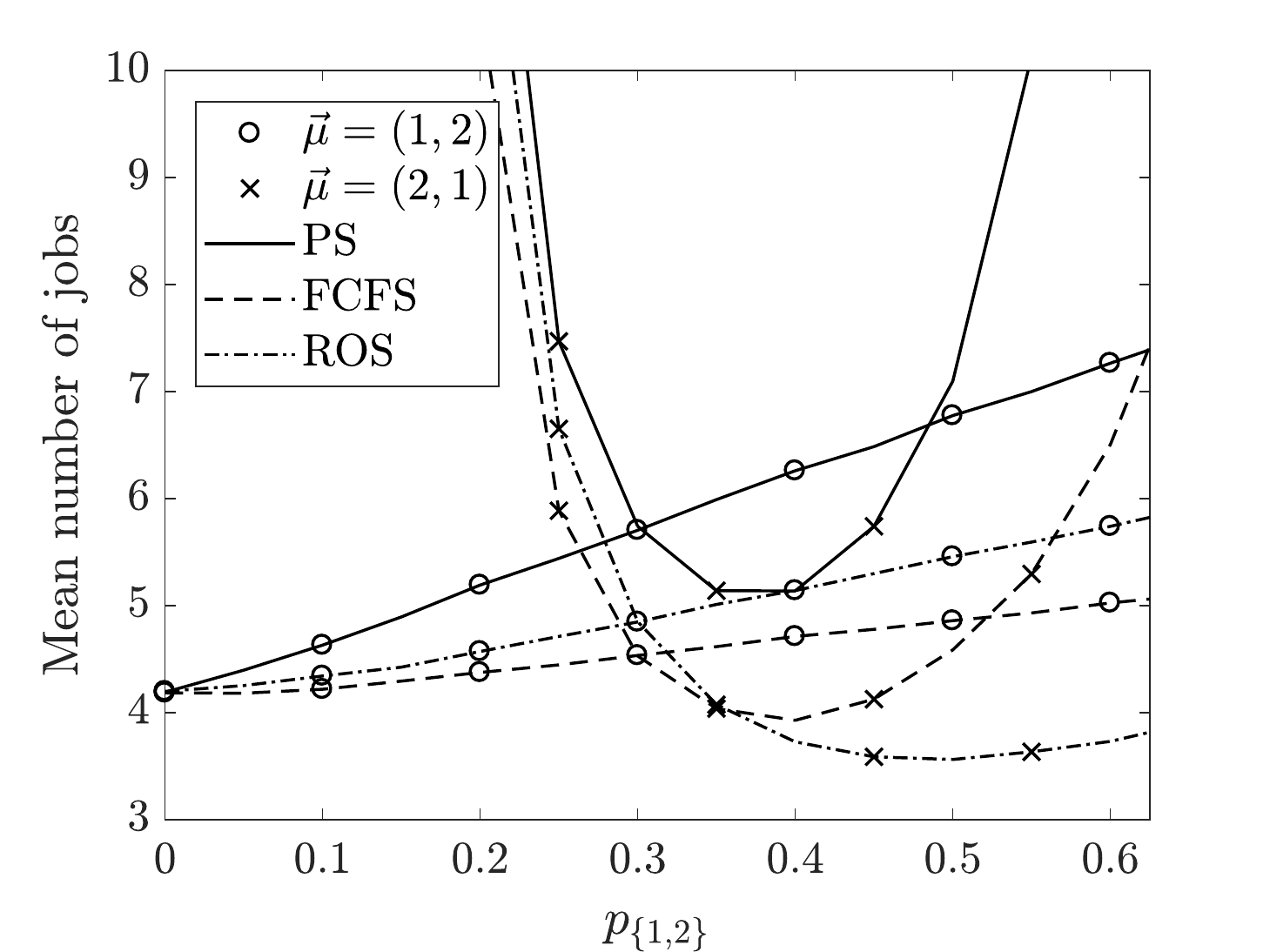}\\
		b) $\lambda=2$ 
	\end{subfigure}
	\begin{subfigure}{.32\textwidth}
		\centering
		\includegraphics[width=1.1\textwidth]{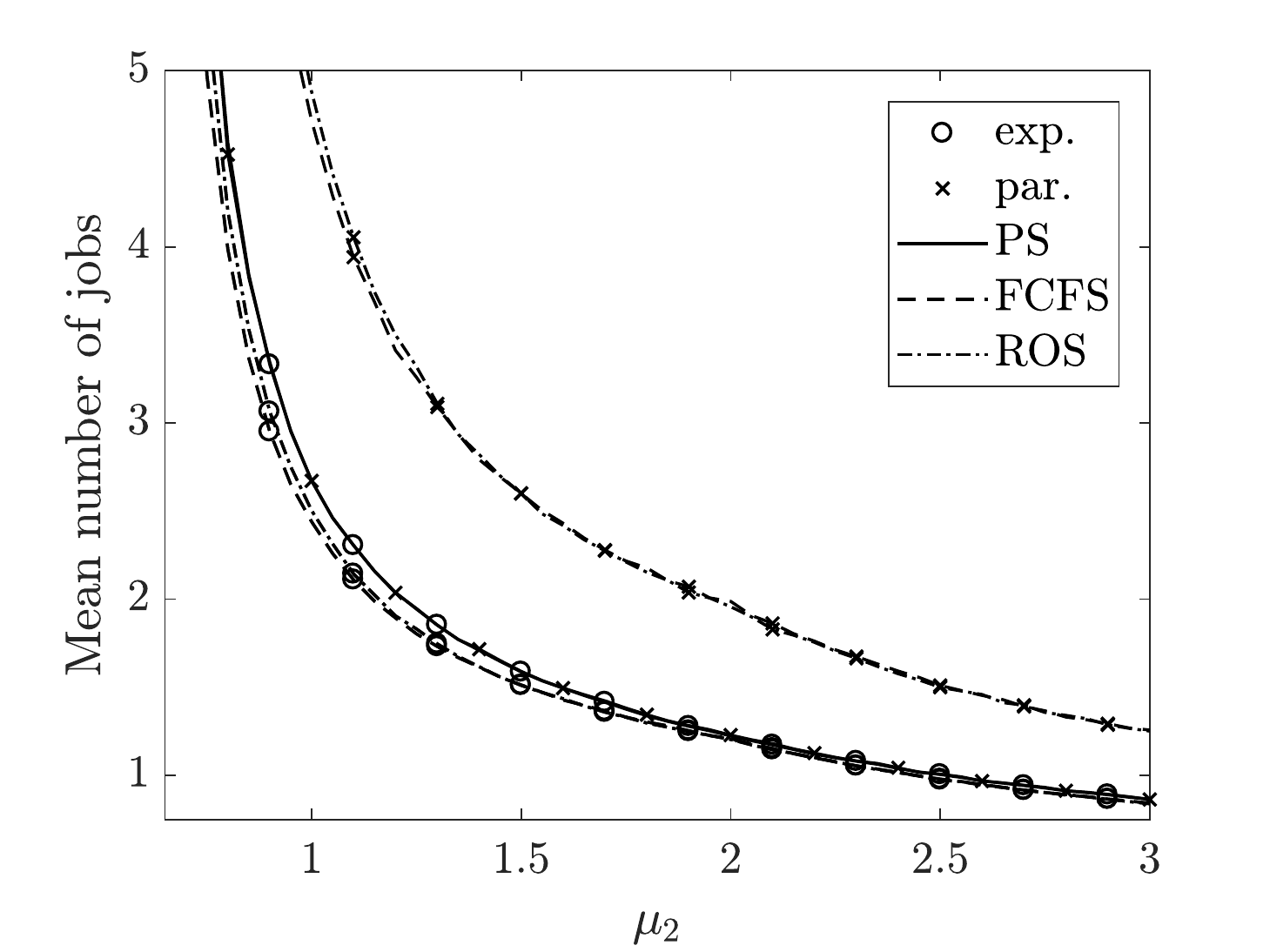}\\
		c) $\vec p = (0.35, 0.40, 0.25)$
	\end{subfigure}
	\caption{Mean number of jobs with redundancy combined with PS, FCFS, and ROS. a) and b) for the $W$ model with respect to $p_{\{1,2\}}$ and exponentially distributed service times, with $p_{\{1\}}=0.35$ and $p_{\{2\}} = 1- p_{\{1\}}-p_{\{1,2\}}$. a) $\lambda =1 $, b) $\lambda = 2$. c) For the $W$ model under exponentially and bounded Pareto ($\alpha=0.5$, $\tilde q=15$) distributed service times, and with respect to $\mu_2$, for $\lambda=1.5$, $\vec p = (0.35, 0.4, 0.25)$ and $\mu_1=2$. }
	\label{Fig:F-R}
\end{figure}

In Figure~\ref{Fig:F-R} a) and b) we consider a $W$-model and compare the performance  for the different policies  PS, FCFS and ROS. We take  exponentially distributed service times. We plot the mean number of  jobs with  respect to $p_{\{1,2\}}$, with $p_{\{1\}}=0.35$ and $p_{\{2\}}=1-p_{\{1\}}-p_{\{1,2\}}$. 
In Figure~\ref{Fig:F-R}~a) we set $\lambda=1$, and in Figure~\ref{Fig:F-R}~b) we set  $\lambda=2$. The stability condition under PS is given in Figure~\ref{Fig:W_exp1} c).

In the case of $\vec \mu=(1,2)$, we observe that FCFS always outperforms ROS. Intuitively we can explain this  as follows. Since $p_{\{1\}}$ is kept fixed, as $p_{\{1,2\}}$ increases,  the load in server~1 increases. With FCFS, it is more likely that both servers work on the same copy, and hence that  the fast server~2 ``helps'' the slow server 1 (with high load). With ROS however, both servers tend to work on different copies, and the loaded slow server~1 will  take a long time serving copies that could have been served faster in the fast server~2.  
On the other hand, with $\vec\mu=(2,1)$ and sufficiently large $p_{\{1,2\}}$, ROS outperforms FCFS. In this case, the loaded server 1 is the fast server, and hence having both servers working on the same copy becomes ineffecient, which explains that the performance under ROS becomes better. As a  rule of thumb, it seems that for a redundancy model,  if slow servers are highly loaded, then FCFS is preferable, but if fast servers are highly loaded, then ROS is preferable. 

From Figures~\ref{Fig:F-R} a) and b) we further observe that for all values of $p_{\{1,2\}}$, FCFS and ROS outperform PS, and that the gap increases when $\lambda$ increases. 
In Figure~\ref{Fig:F-R} c) we consider exponential and bounded Pareto (with $\alpha=0.5$ and $\tilde q=15$) service time distributions and plot the mean number of jobs  for different values of $\mu_2$, when $\lambda=1.5$, $\vec p=(0.35, 0.4, 0.25)$ and $\mu_1=2$. As before, with exponentially distributed service times, FCFS and ROS slightly outperform PS. In the case where jobs have bounded Pareto distributed service times, PS outperforms both FCFS and ROS. This seems to indicate that as the variability of the service time distribution increases, PS might become a preferable choice over FCFS and ROS in redundancy systems. Additionally, under PS we observe that the mean number of jobs is nearly insensitive to the service time distribution.

The main insight we obtain from Figure~\ref{Fig:F-R} is that the stability and performance of heterogeneous redundancy systems strongly depends on the employed service policy in the servers. We leave the stability analysis of other scheduling policies (such as FCFS or ROS) for future work as they require a different proof approach.

\section{Conclusion}
\label{sec:conclusion}
With exponentially distributed jobs, and i.i.d. copies, it has been shown that redundancy does not reduce the stability region of a system, and that it improves the performance. This happens in spite of the fact that redundancy necessarily implies a waste of computation resources in servers that work on copies that are canceled before being fully served. The modeling assumptions play thus a crucial role, and as argued in several papers, e.g.\ \cite{Gardner17b}, the i.i.d. assumption might lead to insights that are qualitatively wrong.

In the present work, we consider the more realistic situation in which copies are identical, and the service times are generally distributed. We have shown that redundancy can help improve the performance in case the servers capacities are sufficiently heterogeneous.  To the best of our knowledge, this is the first positive result on redundancy with identical copies, and it illustrates that the negative result proven in \cite{Anton2019} critically depends on the fact that the capacities were homogeneous.

We thus believe that our work opens the avenue for further research to understand when redundancy is beneficial in other settings. For instance, it would be interesting to investigate what happens in case servers implement other scheduling policies. It is also important to consider other cross-correlation structures for the copies, in particular the $S\&X$ model recently proposed in the literature. Another interesting situation is when the capacities of the servers fluctuate over time. 
Other possible extension is to consider the cancel-on-start variant of redundancy, in which as soon as one copy enters service, all the others are removed. For conciseness purposes, in this paper we have restricted ourselves to what we considered one of the most basic, yet interesting  and relevant setting.

\section*{Acknowledgments}
The PhD project of E. Anton is funded by the French “Agence Nationale de la
Recherche (ANR)” [Project ANR-15-CE25-0004 (ANR JCJC RACON)]. This work (in particular,
research visits of E. Anton and M. Jonckheere) was partially funded by a STIC AMSUD GENE
project. U. Ayesta received funding from the Department of Education of the Basque Government through the Consolidated Research Group MATHMODE (IT1294-19).

\bibliographystyle{plain}

\appendix

\section*{APPENDIX}
\section{Proofs of Section~\ref{sec:gen_red}}

\ \\ \textbf{Proof of Corollary~\ref{stab:cond2}}

Let us consider $s\in\mathcal R$. Let $i$ be such that  $s\in\mathcal L_i$, which is  unique since $\{\mathcal L_i\}_{i=1}^{i^*}$ is a partition of $\mathcal R$. 
We will show that for this $s$ and $i$, it holds that $CAR_i = \frac{\mu_s}{\sum_{c: s\in \mathcal{R}(c)} p_c}$. Hence, together with Corollary~\ref{stab:cond} this concludes the result. 

First, note that $CAR_i = \frac{\mu_s}{\sum_{c\in C_i(s)} p_c}$. Hence, we need to prove that $\sum_{c: s\in \mathcal{R}(c)} p_c = \sum_{c\in C_i(s)} p_c$, or equivalently, $\{{c: s\in \mathcal{R}(c)}\} = C_i(s)$.

For any $c\in\mathcal C(s)$, $\mathcal R(c)=\mathcal L_l(c)$ with $l\leq i$. We note that $\mathcal C_i(s) = \mathcal C(s) \backslash \{c\in\mathcal C(s) \ : \ \mathcal R(c)=\mathcal L_l(c) \textrm{ with } l<i\}$. Therefore,  for $s\in\mathcal L_i$, $\mathcal C_i(s)=\{c\in\mathcal C \ : s\in c \ , \ c\in\mathcal C_i \ , \ s\in\mathcal L_i(c) \} = \{c\in\mathcal C \ :  s\in \mathcal R(c) \}$. The last equality holds by definition of $\mathcal R(c)$. 
\ \hfill $\Box$

\ \\ \textbf{Proof of Corollary~\ref{corol:homo}.}

The stability condition of such a system is given by Corollary~\ref{stab:cond}. We note that each server $s\in S$ receives $\mathcal C(s)=\binom{K-1}{d-1}$ different job types, that is, by fixing a copy in server $s$, all possible combinations of $d-1$ servers out of $K-1$. Thus, $\mathcal L_1=\arg\max_{s\in S_1}\{\frac{\binom{K}{d}}{\binom{K-1}{d-1}}\mu_s\}=K$, $S_2=S-\{K\}$ and condition $\lambda\frac{\binom{K-1}{d-1}}{\binom{K}{d}}<\mu_K$. 

We note each server $s\in S_i$ receives $\binom{\vert S_i\vert-1}{d-1}$ different job types, for $i=1,\ldots,i^*$ and thus, the maximum capacity-to-fraction-of-arrivals ratio in the subsystem with servers $S_i$, only depends on the capacities of servers in $S_i$, that is $\mathcal L_i=\arg\max_{s\in S_i}\{\mu_s\}$. Additionally since, $\mu_1<\ldots<\mu_K$, one obtains that $\mathcal L_i=K-i+1$, for $i=1,\ldots, K-d+1$. The associated conditions are $\lambda\frac{\binom{K-i+1}{d-1}}{\binom{K}{d}}<\mu_{K-i+1}$ for $i=1,\ldots,K-i+1$. This set of conditions is equivalent to that in Corollary~\ref{corol:homo}.  
\ \hfill $\Box$

\section{Proofs of Section~\ref{sec:suf}}

We first introduce some notation: 
We denote by $E_c(t)=\max\{j\ : \ U_{cj}<t \}$ the number of type-$c$ jobs that arrived during the time interval $(0,t)$ and by $U_{cj}$ the instant of time at which the $j$th type-$c$ job arrived to the system. 
We recall that $b_{cj}$ denotes its service realization. 
We denote by $b'_{cms}$ the residual job size of the $m$th eldest type-$c$ job in server $s$ that is already in service at time $0$. 

\subsection*{Sufficient stability condition}

\ \\ \textbf{Proof of Proposition \ref{prop:suff}}

We now prove the stability of the UB system.  For that, we first describe   the dynamics of the number of type-$c$ jobs in the UB system, denoted by $N_c^{UB}(t)$. We recall that a type-$c$ job departs only when all the copies in the set of servers $\mathcal R(c)$ are completely served. 
We let  $\eta_{\mathcal R(c)}^{min}(v,t) = \min_{\tilde s\in\mathcal R(c)} \{\eta_{\tilde s}(v,t) \}$ be the minimum cumulative amount of capacity received by a copy in one of its servers $\mathcal R(c)$ during the interval $(v,t)$. 
Therefore, 
\begin{eqnarray*}
	N^{UB}_c(t)&=& \sum_{m=1}^{N_c^{UB}(0)} 1\left( \{\exists\tilde s\in\mathcal R(c)\ : \ b'_{cm\tilde s}>\eta_{\tilde s}(0,t)\}\right) + \sum_{j=1}^{E_c(t)} 1\left(b_{cj}>\eta_{\mathcal R(c)}^{min}(U_{cj},t) \right).
\end{eqnarray*}



We denote the number of type-$c$ copies in server $s$ by $M_{s,c}^{UB}(t)$. We note that for a type-$c$ job in server $s$ there are two possibilities: 
\begin{itemize}
	\item 
	if $s\in \mathcal R(c)$, 
	the copy of the type-$c$ job leaves the server as soon as it is completely served. The cumulative amount of capacity that the copy receives during $(v,t)$ is $\eta_s(v,t)$. 
	
	\item If $ s \notin \mathcal R(c)$, the copy of the type-$c$ job in server $s$ leaves the system either if it is completely served or if all copies of this type-$c$ job in the servers $\mathcal R(c)$ are served. We note that for any $\tilde s\in\mathcal R(c)$, $\tilde s\in \mathcal L_{l}$, with $l<i$. 
\end{itemize}
Hence, the number of type-$c$ jobs in server $s\in\mathcal L_i$ is given by the following expression. If $s\in\mathcal R(c)$, 
\begin{eqnarray*}
	M_{s,c}^{UB}(t) &=& \sum_{m=1}^{M_{s,c}^{UB}(0)}1\left(b'_{cms}> \eta_s(0,t) \right) 
    +\sum_{j=1}^{E_c(t)} 1\left(  b_{cj}> \eta_s(U_{cj},t) \right)
\end{eqnarray*}
and if $s\notin\mathcal R(c)$, 
\begin{eqnarray}
	M_{s,c}^{UB}(t) & = &\sum_{m=1}^{M_{s,c}^{UB}(0)} 1\left( \{\exists\tilde s\in\mathcal R(c)\ : \ b'_{cm\tilde s}>\eta_{\tilde s}(0,t)\} \cap b'_{cms}> \eta_s(0,t) \right)\nonumber\\
	&&+ \sum_{j=1}^{E_c(t)} 1\left(b_{cj}>\eta_{\mathcal R(c),s}(U_{cj},t)\right),\nonumber
\end{eqnarray}
where $\eta_{\mathcal R(c),s}(v,t) = \max \{\eta_{\mathcal R(c)}^{min}(v,t),\eta_s(v,t) \}$.
The first terms in both equations correspond to the type-$c$ jobs that where already in the system by time $t=0$, the second terms correspond to the type-$c$ jobs that arrived during the time interval $(0,t)$.

In the following we obtain the number of copies per server. 
Before doing so, we need to introduce some additional notation. 
Let $\mathcal D^l(s)=\{c\in\mathcal C(s) \ : \ \mathcal R(c)\subseteq \mathcal L_l(c)\}$ be the set of types in server $s$
for which the set of servers where these types receive maximum capacity-to-fraction-of-arrivals ratio is $\mathcal R(c)\subseteq\mathcal L_l(c)$. If $s\in \mathcal{L}_i$, then, by definition, $\mathcal D^l(s)\neq \emptyset$ if  $l\leq i$ and   $\{\mathcal D^l(s) \}_{l=1}^i$ forms a partition of $\mathcal C(s)$. Furthermore, $D^i(s)=\mathcal C_i(s)$, for  all $s\in \mathcal L_i$. Therefore, for a server $s\in\mathcal L_i$, the number of copies in the server is given by the following expression: 
\begin{equation*}
	M_{s}^{UB}(s) =\sum_{c\in\mathcal C(s)} M_{s,c}^{UB}(t) = \sum_{l=1}^{i-1} \sum_{c\in{\mathcal D}^{l}(s)} M_{s,c}^{UB}(t) +  \sum_{c\in{\mathcal C}_i(s)} M_{s,c}^{UB}(t).
\end{equation*}
The first term of the RHS of the equation corresponds to the type-$c$ jobs in server $s$ that have $\mathcal R(c)\subseteq\mathcal L_l(c)$. The second term of the RHS corresponds to type-$c$ jobs in server $s$ that have $\mathcal R(c)\subseteq \mathcal L_i(c)$. Parti-cularly, we note that in the UB system, $M^{UB}_s(t)\leq \sum_{c\in\mathcal C(s)} N^{UB}_c(t)$, since copies might have left, while the job is still present. 

In order to prove the stability condition, we investigate the fluid-scaled system.  The fluid-scaling consists in studying the rescaled sequence of systems indexed by parameter $r$. For $r>0$, denote by $M_{c,s}^{UB,r}(t)$ the system where the initial state satisfies $M^{UB}_{s,c}(0)=rm^{UB}_{s,c}(0)$, for all $c\in \mathcal C$ and $s\in S$. We define,  
\begin{eqnarray}
\label{eq:frelationiid}
M_{s,c}^{UB,r}(t)= \frac{M^{UB}_{s,c}(rt)}{r}, \ \mbox{ and } \ 
M_s^{UB,r}(t) = \frac{M^{UB}_s(rt)}{r} \nonumber
\end{eqnarray}

In the following, we give the characterization of the fluid model. 
\begin{definition}
	Non-negative continuous functions $m_s^{UB}(\cdot)$ are a fluid model solution if they satisfy the functional equations 
	\begin{eqnarray}\label{eq:fluid}
	m_s^{UB}(t)&=&
	\sum_{l=1}^{i-1} \sum_{c\in{\mathcal D}^{l}(s)} \left[ m^{UB}_{s,c}(0)\left( 1 - G\left(\bar\eta_{\mathcal R(c),s}(0,t)\right) \right)
	+ \lambda p_c \left(\int_{x=0}^t 1 - F\left(\bar\eta_{\mathcal R(c),s}(x,t) \right) dx\right)\right] \nonumber\\
	&& + \sum_{c\in{\mathcal C}_i(s)} \left[  m_{s,c}^{UB}(0) \left(1 - G(\bar\eta_s(0,t))\right)
	+ \lambda p_c \int_{x=0}^t(1- F(\bar\eta_s(x,t)))dx \right],
	\end{eqnarray}	
	for $s\in\mathcal L_i$ and $i=1,\ldots,i^*$, where $G(\cdot)$ is the  distribution of the remaining service requirements, $F(\cdot)$ the service time distribution of arriving jobs, and 
	$$ \bar\eta_s(v,t)=\int_{x=v}^{t} \phi_s(\vec m^{UB}(x)) dx,$$
	$$ \bar\eta_{\mathcal R(c)}^{min}(v,t)=\min_{\tilde s\in\mathcal R(c)} \{\bar\eta_{\tilde s}(v,t)\}, $$
	$$ \bar\eta_{\mathcal R(c),s}(v,t)=\max \{\bar\eta_{\mathcal R(c)}^{min}(v,t),\bar\eta_{s}(v,t)\}.$$
\end{definition}

The existence and convergence of the fluid limit to the fluid model can now be proved. 

\begin{proposition}\label{l:fluidex}
	The limit point of any convergent subsequence of $(\vec M^{UB,r}(t); t\geq 0)$ is almost surely a solution of the fluid model~\eqref{eq:fluid}. 
\end{proposition}

\textit{Proof of Proposition~\ref{l:fluidex}} The proof is identical to the the proof of Theorem 5.2.1 in \cite{Egorova2009} ({which is itself based on Lemma 5 in \cite{Gromoll2008}}). We only need to ensure that $\bar\eta^{min}_{\mathcal R(c)}(v,t)$ and $ \bar\eta_{\mathcal R(c),s}(v,t)$   are decreasing in $v$ and continuous on $v\in[\psi_s(t)+\epsilon,t]$, where $\psi_s(t)=\sup(v\in[0,t]\ : \ m_s(u)=0)$.

Let us verify that $\bar\eta^{min}_{\mathcal R(c)}(v,t)$ and $ \bar\eta_{\mathcal R(c),s}(v,t)$ are decreasing and continuous on $v$.  We note that the function $\eta_s(\cdot,t)$ that gives the cumulative service that a copy in server~$s$ received during time interval $(\cdot,t)$, is a Lipschitz continuous function, increasing  
for $t<\tau_s$ and non decreasing for $t>\tau_s$, where    $\tau_s=\inf\{t>0 \ : \ M_s(t)=0\}$.

If $\bar\eta^{min}_{\mathcal R(c)}(v,t)= \bar\eta_{s_1}(v,t)$ and $\bar\eta_{\mathcal R(c),s}(v,t)=\bar\eta_{s_2}(v,t)$ for all $v\in[0,t)$ and some $s_1,s_2\in S$, then both $\bar\eta^{min}_{\mathcal R(c)}(v,t)$ and $ \bar\eta_{\mathcal R(c),s}(v,t)$ are decreasing and continuous on $v$, since by definition $\bar\eta_{s}(v,t)$ is decreasing and continuous on $v$ for all $s\in S$. 

Let us assume that for $v_0\in[0,t)$ is such that  $\bar\eta^{min}_{\mathcal R(c)}(v,t)=\bar\eta_{\tilde s^1}(v,t)$ for $v\leq v_0$ and  $\bar\eta^{min}_{\mathcal R(c)}(v_0^+,t)=\bar\eta_{\tilde s^2}(v_0,t)$, for some $\tilde s^1,\tilde s^2\in\mathcal R(c)$. We first verify that $\bar\eta^{min}_{\mathcal R(c)}(v,t)$ is continuous on $v=v_0$. Since, $\bar\eta_{\tilde s^1}(v,t)$ and $\bar\eta_{\tilde s^2}(v,t)$ are continuous on $v=v_0$, then 
$$ \lim\limits_{x^-\to v_0}\bar\eta^{min}_{\mathcal R(c)}(x,t) = \bar\eta_{\tilde s^1}(v,t) = \bar\eta_{\tilde s^2}(v,t)=\lim\limits_{x^+\to v_0}\bar\eta^{min}_{\mathcal R(c)}(x,t).$$
Therefore, we conclude that $\bar\eta^{min}_{\mathcal R(c)}(x,t)$ is continuous on $v\in[0,t)$. Analogously, one can verify that $\bar\eta^{min}_{\mathcal R(c),s}(x,t)$ is continuous on $v\in[0,t)$. 

We now verify that $\bar\eta^{min}_{\mathcal R(c)}(x,t)$ is decreasing on $v\in[0,t)$. Let us consider $0<t_1<v_0<t_2<t$. Then for $\bar\eta^{min}_{\mathcal R(c)}(v,t)$,
$$ \bar\eta^{min}_{\mathcal R(c)}(t_1,t)= \bar\eta_{\tilde s^1}(t_1,t) \leq \bar\eta_{\tilde s^1}(t_2,t) \leq \bar\eta_{\tilde s^2}(t_2,t)=\bar\eta^{min}_{\mathcal R(c)}(t_2,t),$$
where the first inequality holds since $\bar\eta_{\tilde s^1}(v,t)$ is decreasing on $v$. We conclude that $\bar\eta^{min}_{\mathcal R(c)}(v,t)$ is decreasing $v$. 

Let us verify that $ \bar\eta_{\mathcal R(c),s}(v,t)$ is decreasing on $v$. W.l.o.g. we assume that there exists $v_0\in[0,t)$, such that $\bar\eta_{\mathcal R(c),s}(v,t)= \bar\eta^{min}_{\mathcal R(c)}(v,t)$ for $v<v_0$ and $\bar\eta_{\mathcal R(c),s}(v,t)= \bar\eta_{s}(v,t)$ for $t>v>v_0$. Then, 
$$ \bar\eta_{\mathcal R(c),s}(t_1,t)= \bar\eta^{min}_{\mathcal R(c)}(t_1,t) \leq \bar\eta_{s}(t_1,t)\leq \bar\eta_{s}(t_2,t)=\bar\eta_{\mathcal R(c),s}(t_2,t)$$
where the first inequality  holds since $\bar\eta_{\tilde s^1}(v,t)$ is decreasing on $v$. We conclude that $\bar\eta_{\mathcal R(c),s}(x,t)$ is decreasing $v$. 
\ \hfill $\Box$
\bigskip

We now give a further characterization of the fluid model~\eqref{eq:fluid}. 

\begin{proposition}
	\label{prop:fluidlimit}
	Let $i\leq i^*$ and assume  $\lambda\sum_{c\in{\mathcal C}_l( s)} p_c < \mu_{s}$ for all $l\leq i-1$ and $s\in  \mathcal L_l$.
	Then, there is a time $T\geq 0$, such that for $t\geq T$ and for  $s\in \cup_{l=1}^{i-1}\mathcal L_l$, 
	$
	m_s^{UB}(t)=0
	$
	and for $s\in \mathcal L_i$
	\begin{eqnarray}\label{eq:fluid2}
	m_s^{UB}(t)&=&\sum_{c\in{\mathcal C}_i(s)} \left[ m_{s,c}^{UB}(0) \left(1 - G(\bar\eta_s(0,t))\right)
	+ \lambda p_c \int_{x=0}^t(1- F(\bar\eta_{s,i}(x,t)))dx \right],
	\end{eqnarray}	
	with
	$$ \bar\eta_{s,i}(v,t):=\int_{x=v}^{t} \phi_{s,i}(\vec m(x)) dx,$$
	and $\phi_{s,i}(\vec m(x)):= \frac{\mu_s}{\sum_{c\in\mathcal C_i(s)} m_{s,c}(x)}$. 
\end{proposition}

\textit{Proof of Proposition \ref{prop:fluidlimit}} For simplicity in notation, we remove the superscript $UB$ throughout the proof. 

First assume $s\in \mathcal{L}_1$.  Since $\mathcal{D}^0=\emptyset$, from Equation~\eqref{eq:fluid}, we directly obtain 
\begin{eqnarray}
m_s(t)&=&\sum_{c\in\mathcal C_1(s)} [ m_{s,c}(0) \left(1 - G(\bar\eta_s(0,t))\right) +\lambda p_c \int_{x=0}^t(1- F(\bar\eta_s(x,t)))dx]\ , \ \ \forall t>0. \nonumber
\end{eqnarray} 
This expression coincides with the fluid limit of an $M/G/1$ PS queue with arrival rate \\
$\lambda \sum_{c\in\mathcal C_1(s)} p_c$ and server speed $\mu_s$. Since $\lambda \sum_{c\in\mathcal C_1(s)} p_c$ $<\mu_s$, we know that there exists a  $\bar \tau_s$ such that  $m_{s}(t)=0$, for all $t\geq \bar \tau_s$. 

The remainder of the proof is by induction. Consider now a server $s\in \mathcal{L}_l$ and assume there exists a time $\tilde T$ such that $m_{s}(t)=0$, for all $t\geq \tilde T$ and $s\in \cup_{j=1}^{l-1}\mathcal{L}_j$.  
Thus, for $t\geq\tilde T$, also $m_{s,c}(t)=0$ for all $s\in \cup_{j=1}^{l-1}\mathcal{L}_j$, $c\in\mathcal D^j(s)$, $j=1,\ldots,l-1$. 
We consider server $s\in\mathcal{L}_l$.
From~\eqref{eq:fluid} its drift is then given by: 
\begin{eqnarray}
m_s(t)
&=&\sum_{j=1}^{l-1} \sum_{c\in{\mathcal D}^{j}(s)} m_{s,c}(t)+ \sum_{c\in{\mathcal C}_l(s)} m_{s,c}(t)\nonumber \\
& = &\sum_{c\in{\mathcal C}_l(s)} \left[m_{s,c}(0) \left(1 - G(\bar\eta_s(0,t))\right)
+\lambda p_c\int_{x=0}^t (1- F(\bar\eta_s(x,t)))dx\right],\nonumber
\end{eqnarray}
for all $t\geq \tilde T$. 
Now note that $\phi_s(\vec m(t))=\frac{\mu_s}{  m_{s}(t)}=\frac{\mu_s}{\sum_{c\in{\mathcal C}_l(s)} m_{s,c}(t)}= \phi_{s,l}(\vec m(t))$, where the second equality follows from the fact that  $m_{s,c}(t)=0$  for all for all $s\in \cup_{j=1}^{l-1}\mathcal{L}_j$, $c\in\mathcal D^j(s)$, $j=1,\ldots,l-1$. 

To finish the proof, \eqref{eq:fluid2} coincides with the fluid limit of an $M/G/1$ system with PS, arrival rate $\lambda \sum_{c\in{\mathcal C}_l(s)} p_c$ and server speed $\mu_s$. Hence, if $l<i$,  the standard PS queue is stable,  and we are sure that it equals and remains zero in finite time.
\ \hfill $\Box$
\bigskip

Below we prove that the $UB$ system is Harris recurrent. Note that the concept of Harris recurrence is needed here since the state space is obviously not countable, (as we need to keep track of residual service times).
We first establish the fluid stability,  that is, the fluid model  is $0$ in finite time. The latter is useful, as we can use  the results 
of \cite{Lee2008} that establish that under some suitable conditions, fluid stability implies Harris recurrency, see the lemma below.


\begin{lemma}\label{lemma:Harris}
	If the fluid limit    is fluid stable, then the  stochastic system is Harris recurrent.
\end{lemma}

\textit{Proof of Lemma~\ref{lemma:Harris}} In \cite{Lee2008}, the authors consider bandwidth sharing networks (with processor sharing policies), and show that under mild conditions, the stability of the fluid model (describing the Markov process of the number of per-class customers with their residual job sizes) is sufficient for stability (positive Harris recurrence).


Our system, though slightly different from theirs satisfies   the same assumptions, and as a consequence their results are directly applicable to our model. 


More precisely, given the assumptions on the 
service time distribution, our model satisfies the assumptions given in \cite[Section 2.2]{Lee2008} for inter-arrival times and job-sizes. (In particular exponential inter-arrival times satisfy the conditions given in \cite[Assumption 2.2.2]{Lee2008}.) 
%
\ \hfill $\Box$
\bigskip

Equation~\eqref{eq:fluid2} coincides with the fluid limit of an $M/G/1$ PS system with arrival rate \\
$\lambda \sum_{c\in{\mathcal C}_i(s)} p_c$ and server speed $\mu_s$.
If $\lambda < CAR _l,$  or equivalently $\lambda \sum_{c\in{\mathcal C}_i(s)} p_c< \mu_s$,   for all  $l=1,\ldots,i,$ Equation~\eqref{eq:fluid2} equals zero in finite time.
Hence,  from Lemma~\ref{lemma:Harris} we  conclude that for servers $s\in\mathcal L_i$, the associated stochastic number of copies in server~$s$ is Harris recurrent, as stated in the corollary below. 
\ \hfill $\Box$

\ \\ \textbf{Proof of Proposition \ref{prop:ub}}

We assume that both systems are coupled as follows: at time $t=0$, both systems start at the same initial state $N_c(0)=N^{UB}_c(0)$ and $a_{cjs}(0)=a^{UB}_{cjs}(0)$ for all $c,j,s$. Arrivals and service times are also coupled. For simplicity in notation, we assume that when in the original system a type-$c$ copy reaches its service requirement $b$, the attained service of its $d-1$ additional copies is fixed to $b$ and the job remains in the system until the copy of that same job in the $UB$ system is fully served at all servers in $\mathcal R(c)$.  

We prove this result by induction on $t$. It holds at time $t=0$. We assume that for $u\leq t$ it holds that $N_c(t)\leq N^{UB}_c(t)$ and $a_{cjs}(t)\geq a^{UB}_{cjs}(t)$ for all $c,j,s$. We show that this inequality holds for $t^+$. 

We first assume that at time $t$, it holds that $N_c(t)=N_c^{UB}(t)$ for some $c\in \mathcal C$. The inequality is violated only if there is a job for which the copy in the $UB$ system is fully served at all servers $\mathcal R(c)$, but none of the  copies in the original system is completed. That means, there exist a $j$ such that $a_{cjs}(t)<a_{cj\mathcal R(c)}^{UB}(t)=b_j$ for all $s\in c$. However, this can not happen, since by hypothesis $a_{cjs}(t)\geq a^{UB}_{cjs}(t)$ for all $s\in c$.

We now assume that at time $t$, $a_{cjs}(t)=a^{UB}_{cjs}(t)$ for some $c,j,s$. There are now two cases.
If this copy (and job) has already left in the original system, then $a_{cjs}(t)=a_{cjs}(t^+)=b_{cj}$ and hence $a_{cjs}(t^+)\geq a^{UB}_{cjs}(t^+)$. If instead the copy has not left in the original system, then by hypothesis it holds that $N_c(t)\leq N_c^{UB}(t)$ and thus, $M_s(t)\leq M_s^{UB}(t)$ and $\frac{\mu_s}{M_s(t)}\geq \frac{\mu_s}{M_s^{UB}(t)}$. That means that the copy in the original system has a higher service rate at time $t$ than the same copy in the $UB$ system. 
Hence,  $a_{cjs}(t^+)\leq a^{UB}_{cjs}(t^+)$. 
\ \hfill $\Box$

\subsection*{Necessary stability condition}

\ \\ \textbf{Proof of Proposition~\ref{corol:LB}}

In order to show that the $LB$ system is unstable, we investigate the fluid-scaled system.  For $r>0$, denote by $N_{c}^{LB,r}(t)$ the system where the initial state satisfies $N^{LB}_{c}(0)=rn^{LB}_{c}(0)$, for all $c\in \mathcal C$. We write for the fluid-scaled number of jobs per type 
\begin{eqnarray}
\label{eq:fluLB}
&& N_c^{LB,r}(t)= \frac{N^{LB}_c(rt)}{r}. \nonumber
\end{eqnarray}
In the following we give the characterization of the fluid model.
\begin{definition}
	\label{def:3}
	Non-negative continuous functions $n_c^{LB}(\cdot)$ are a fluid model solution if they satisfy the functional equations 
	\begin{eqnarray}\label{eq:fluidLB}
	&&n_c^{LB}(t)= 0, \ \ c\in\mathcal{C}\backslash\mathcal{C}_\iota\nonumber\\
	&&n_c^{LB}(t)= n^{LB}_{c}(0)\left( 1 - G\left(\bar\eta^{LB}_{c}(0,t)\right) \right) 
	+ \lambda p_c \left(\int_{x=0}^t 1 - F\left(\bar\eta^{LB}_{c}(x,t) \right) dx\right), \ \ c\in\mathcal C_\iota\nonumber
	\end{eqnarray}	
	where $G(\cdot)$ is the  distribution of the remaining service requirements of initial jobs, $F(\cdot)$ the service time distribution of arriving jobs and
	$$ \bar\eta^{LB}_c(v,t)=\int_{x=v}^{t} \phi^{LB}_c(\vec n^{LB}(x)) dx, \ \mbox{ 	with } \ c\in\mathcal C_\iota.$$
\end{definition}

The existence and convergence of the fluid-scaled number of jobs $\vec N^{LB,r}(t)$ to the fluid model $\vec n^{LB}(t)$ can be proved as before. The statement of Proposition~\ref{l:fluidex}, indeed directly translates to the process $\vec N^{LB,r}(t)$, since $\eta^{LB}_{c}(v,t)$ is both decreasing and continuous in $v$. Therefore, it is left out.




Next, we characterize the fluid model solution $\vec n^{LB(t)}$ in terms of $m_s^{LB}(t)=\sum_{c\in \mathcal{C}(s)}n^{LB}_c(t)$. We  show that if the initial condition for all servers is such that  $m^{LB}_s(0)/\mu_s^{LB}=\alpha(0)$ for all $s\in S_\iota$, then     $m^{LB}_s(t)/\mu_s^{LB}=\alpha(t)$ for all $s\in S_\iota$, where $\alpha(t)$ is given below.

\begin{lemma}\label{lemma:LB2}
	Let us assume that the initial condition is such that $n^{LB}_c(0)=0$ for all $c\in\mathcal C\backslash\mathcal C_\iota$ and for $c\in\mathcal C_\iota$, $n^{LB}_c(0)$ are such that $m^{LB}_s(0)/\mu_s^{LB}=\alpha(0)$ for all $s\in S_\iota$.		
	Let
	\begin{equation}\label{eq:2}
	\alpha(t)=\alpha(0)(1-G(\bar\eta^{LB}_\alpha(0,t))) + \frac{\lambda}{CAR_\iota}\int_{x=0}^t(1-F(\bar\eta^{LB}_\alpha(x,t))) dx,
	\end{equation}
	where $\bar\eta^{LB}_\alpha (v,t)= \int_{x=v}^t \phi^{LB}_\alpha(\alpha(x))dx$,
	with $\phi^{LB}_\alpha(\alpha(t))=\frac{1}{\alpha(t)}$.
	
	Then, $n^{LB}_c(t)=0$ for all $t\geq 0$ and $c\in\mathcal C\backslash\mathcal C_\iota$, and $$m^{LB}_s(t)/\mu_s^{LB}=\alpha(t),$$ for all $t\geq 0$ and $s\in S_\iota$. 
\end{lemma}

\textit{Proof of Lemma \ref{lemma:LB2}} From Definition~\ref{def:3}, we obtain that for each server $s\in S_\iota$,
\begin{eqnarray*}
	\frac{m^{LB}_s(t)}{\mu_s^{LB}}&=& \frac{1}{\mu_s^{LB}}\sum_{c\in{\mathcal C}_{\iota}(s)}  n^{LB}_c(t)\\ 
	&=& \sum_{c\in{\mathcal C}_{\iota}(s)} \left[\frac{n^{LB}_{c}(0)}{\mu_s^{LB}}\left( 1 - G\left(\bar\eta^{LB}_{c}(0,t)\right) \right) 
	+ \frac{\lambda p_c}{\mu_s^{LB}} \left(\int_{x=0}^t 1 - F\left(\bar\eta^{LB}_{c}(x,t) \right) dx\right)\right].
\end{eqnarray*}
We recall that $\alpha(t)$ is defined as 
$$\alpha(t) = \alpha(0) \left( 1 - G\left(\bar\eta^{LB}_{\alpha}(0,t)\right) \right)  + \frac{\lambda}{CAR_\iota} \left(\int_{x=0}^t 1 - F\left(\bar\eta^{LB}_{\alpha}(x,t) \right) dx\right).
$$
We let the initial condition be such that $\frac{m_s^{LB}(0)}{\mu_s^{LB}}=\alpha(0)$ for all $s\in S_\iota$ and we will  prove by contradiction that for all $t>0$,	
\begin{align*}\label{eq:3}
	& 	\frac{m^{LB}_s(t)}{\mu_s^{LB}}=\alpha(t), \ \mbox{ for all $s\in S_\iota$.}
\end{align*}
Let us assume that $t_0$ is the first time such that there exists $\tilde s\in S_\iota$ such that $\alpha(t_0)\neq m^{LB}_{\tilde s}(t_0)/\mu_{\tilde s}^{LB}$. 
Since $\sum_{c\in{\mathcal C}_{\iota}(s)} \frac{n^{LB}_{c}(0)}{\mu_s^{LB}} = \frac{m_s^{LB}(0)}{\mu_s^{LB}}=\alpha(0)$ and $\sum_{c\in{\mathcal C}_{\iota}(s)} \frac{p_c}{\mu_s^{LB}} = 1/CAR_\iota,$ this implies that there exist $\tilde c\in\tilde{\mathcal C}$ and $t_1$, $0\leq v\leq t_1<t_0$ such that $\bar\eta^{LB}_\alpha(v,t_1) \neq \bar\eta^{LB}_{\tilde c}(v,t_1)$. 
However, since $\alpha(t)=m^{LB}_s(t)/\mu_s^{LB}$ for all $t<t_0$, this implies that $\phi_c(\vec n(t))=1/\alpha(t)$ for all $t<t_0$ and $c\in\tilde{\mathcal C}(s)$, and hence $\bar\eta^{LB}_\alpha(v,t) = \bar\eta^{LB}_{\tilde c}(v,t)$, for all $t<t_0$. We have hence reached a contradiction, which concludes the proof. 
\ \hfill $\Box$
\bigskip



We note that Equation~\eqref{eq:2} corresponds to the fluid limit of an $M/G/1$ system with PS, arrival rate $\lambda/CAR_\iota$ and server speed $1$. Assuming  $\lambda > CAR_\iota$, it follows that the fluid limit $\alpha(t)$, and hence $m_s^{LB}(t), s\in S_\iota$,  diverges. Now, by using similar arguments as in Dai~\cite{Dai1996}, the fact that the limit diverges implies that the correspon-ding stochastic process can not be tight, and hence cannot be stable.
\ \hfill $\Box$

\ \\ \textbf{Proof of Proposition~\ref{prop1}}

The number of type-$c$ jobs in the system is given by
$N_{c}^{LB}(t) =0,$ for $c\in\mathcal C\backslash\mathcal C_\iota$, and for $c\in\mathcal C_\iota$,
\begin{equation*}
	N_{c}^{LB}(t) =\left[ \sum_{m=1}^{N_{c}^{LB}(0)}1\left(b'_{cms}> \eta^{LB}_c(0,t) \right) +\sum_{j=1}^{E_c(t)} 1\left(b_{cj}> \eta^{LB}_c(U_{cj},t) \right)\right].
\end{equation*}

We note that for all $c\in\mathcal C\backslash\mathcal C_\iota$ the result is direct since $p^{LB}_c=0$ for all  $c\in\mathcal C\backslash\mathcal C_\iota$. Then, let us consider $c\in\mathcal C_\iota$. For any $\vec N$ and $\vec N^{LB}$ such that $\vec N\geq\vec N^{LB}$, the following inequalities hold:

\begin{eqnarray}
\phi_s(\vec N)&=&\frac{\mu_s}{M_s} 
= \frac{\mu_{s}(\sum_{c\in\mathcal C_\iota(s)} p_c)}{(\sum_{c\in\mathcal C_\iota(s)} p_c)M_{s}} =\frac{(\sum_{c\in\mathcal C_\iota(s)} p_c)\mu_{s}/(\sum_{c\in\mathcal C_\iota(s)} p_c)}{\sum_{c\in\mathcal C(s)\backslash\mathcal C_\iota(s)}  N_c+\sum_{c\in\mathcal C_\iota(s)} N_c}\nonumber\\
&\leq&\frac{(\sum_{c\in\mathcal C_\iota(s)} p_c)\mu_{s}/(\sum_{c\in\mathcal C_\iota(s)} p_c)}{\sum_{c\in\mathcal C_\iota(s)} N_c}
\leq \frac{CAR_\iota\times (\sum_{c\in\mathcal C_\iota(s)} p_c)}{\sum_{c\in\mathcal C_\iota(s)} N_c^{LB}}\leq \max_{s\in c}\left\{\frac{\mu_s^{LB}}{M^{LB}_s}\right\}\nonumber\\
&=&\phi^{LB}_{c}(\vec N^{LB}).\nonumber
\end{eqnarray}
The second last inequality holds since $CAR_\iota\geq\mu_{s}/(\sum_{c\in\mathcal C_\iota(s)}p_c)$ for all $s\in S_\iota$ and $N^{LB}_c\leq N_c$ for all $c\in\mathcal C_\iota$. We note that $\sum_{c\in\mathcal C_\iota(s)} N_c^{LB}=M_s^{LB}(t) $. It follows from straight forward sample-path arguments that $N^{LB}_c(t)\leq N_c(t)$ for all $t\geq0$ and $c\in\mathcal C_\iota$.
\ \hfill $\Box$

\end{document}